\documentclass{aastex62}

\usepackage{hyperref}
\usepackage{multirow}
\usepackage{amsmath}
\usepackage{graphicx} 
\usepackage{gensymb}
\usepackage{rotating}
\usepackage{float}
\usepackage{threeparttable, booktabs}

\shorttitle{Inferring Vector Magnetic Fields from Stokes Profiles of GST/NIRIS}
\shortauthors{Liu et al.}

\begin{document}

\title{{\bf \large Inferring Vector Magnetic Fields from Stokes Profiles of GST/NIRIS 
		Using a Convolutional Neural Network}}

\author{Hao Liu}
\affiliation{Institute for Space Weather Sciences, New Jersey Institute of Technology, University Heights, Newark, NJ 07102-1982, USA 
hl422@njit.edu, yan.xu@njit.edu, jw438@njit.edu, ju.jing@njit.edu, chang.liu@njit.edu, wangj@njit.edu, haimin.wang@njit.edu}
\affiliation{Department of Computer Science, New Jersey Institute of Technology, University Heights, Newark, NJ 07102-1982, USA}

\author{Yan Xu}
\affiliation{Institute for Space Weather Sciences, New Jersey Institute of Technology, University Heights, Newark, NJ 07102-1982, USA 
hl422@njit.edu, yan.xu@njit.edu, jw438@njit.edu, ju.jing@njit.edu, chang.liu@njit.edu, wangj@njit.edu, haimin.wang@njit.edu}
\affiliation{Center for Solar-Terrestrial Research, New Jersey Institute of Technology, University Heights, Newark, NJ 07102-1982, USA}
\affiliation{Big Bear Solar Observatory, New Jersey Institute of Technology, 40386 North Shore Lane, Big Bear City, CA 92314-9672, USA}

\author{Jiasheng Wang}
\affiliation{Institute for Space Weather Sciences, New Jersey Institute of Technology, University Heights, Newark, NJ 07102-1982, USA 
hl422@njit.edu, yan.xu@njit.edu, jw438@njit.edu, ju.jing@njit.edu, chang.liu@njit.edu, wangj@njit.edu, haimin.wang@njit.edu}
\affiliation{Center for Solar-Terrestrial Research, New Jersey Institute of Technology, University Heights, Newark, NJ 07102-1982, USA}
\affiliation{Big Bear Solar Observatory, New Jersey Institute of Technology, 40386 North Shore Lane, Big Bear City, CA 92314-9672, USA}

\author{Ju Jing}
\affiliation{Institute for Space Weather Sciences, New Jersey Institute of Technology, University Heights, Newark, NJ 07102-1982, USA 
hl422@njit.edu, yan.xu@njit.edu, jw438@njit.edu, ju.jing@njit.edu, chang.liu@njit.edu, wangj@njit.edu, haimin.wang@njit.edu}
\affiliation{Center for Solar-Terrestrial Research, New Jersey Institute of Technology, University Heights, Newark, NJ 07102-1982, USA}
\affiliation{Big Bear Solar Observatory, New Jersey Institute of Technology, 40386 North Shore Lane, Big Bear City, CA 92314-9672, USA}

\author{Chang Liu}
\affiliation{Institute for Space Weather Sciences, New Jersey Institute of Technology, University Heights, Newark, NJ 07102-1982, USA 
hl422@njit.edu, yan.xu@njit.edu, jw438@njit.edu, ju.jing@njit.edu, chang.liu@njit.edu, wangj@njit.edu, haimin.wang@njit.edu}
\affiliation{Center for Solar-Terrestrial Research, New Jersey Institute of Technology, University Heights, Newark, NJ 07102-1982, USA}
\affiliation{Big Bear Solar Observatory, New Jersey Institute of Technology, 40386 North Shore Lane, Big Bear City, CA 92314-9672, USA}

\author{Jason T. L. Wang}
\affiliation{Institute for Space Weather Sciences, New Jersey Institute of Technology, University Heights, Newark, NJ 07102-1982, USA 
hl422@njit.edu, yan.xu@njit.edu, jw438@njit.edu, ju.jing@njit.edu, chang.liu@njit.edu, wangj@njit.edu, haimin.wang@njit.edu}
\affiliation{Department of Computer Science, New Jersey Institute of Technology, University Heights, Newark, NJ 07102-1982, USA}

\author{Haimin Wang}
\affiliation{Institute for Space Weather Sciences, New Jersey Institute of Technology, University Heights, Newark, NJ 07102-1982, USA 
hl422@njit.edu, yan.xu@njit.edu, jw438@njit.edu, ju.jing@njit.edu, chang.liu@njit.edu, wangj@njit.edu, haimin.wang@njit.edu}
\affiliation{Center for Solar-Terrestrial Research, New Jersey Institute of Technology, University Heights, Newark, NJ 07102-1982, USA}
\affiliation{Big Bear Solar Observatory, New Jersey Institute of Technology, 40386 North Shore Lane, Big Bear City, CA 92314-9672, USA}

\begin{abstract}
We propose a new machine learning approach to Stokes inversion based on a convolutional neural network (CNN) and the 
Milne-Eddington (ME) method.
The Stokes measurements used in this study were taken by the Near InfraRed Imaging Spectropolarimeter (NIRIS) 
on the 1.6 m Goode Solar Telescope (GST) at the Big Bear Solar Observatory. 
By learning the latent patterns in the training data prepared by the physics-based ME tool,
the proposed CNN method is able to infer vector magnetic fields from the Stokes profiles of GST/NIRIS.
Experimental results show that 
our CNN method produces smoother and cleaner magnetic maps than the widely used ME method.
Furthermore, the CNN method is 4$\sim$6 times faster than the ME method, 
and is able to produce vector magnetic fields in near real-time,
which is essential to space weather forecasting.
Specifically, it takes $\sim$50 seconds for the CNN method to process an image of 720$\times$720 pixels 
comprising Stokes profiles of GST/NIRIS.
Finally, the CNN-inferred results are highly correlated to the ME-calculated results and are closer to the ME's results 
with the Pearson product-moment correlation coefficient (PPMCC) being closer to 1 on average
than those from other machine learning algorithms
such as multiple support vector regression and multilayer perceptrons (MLP). 
In particular, the CNN method outperforms the current best machine learning method (MLP) by 2.6\% on average
in PPMCC according to our experimental study.
Thus, the proposed physics-assisted deep learning-based CNN tool 
can be considered as an alternative, efficient method 
for Stokes inversion for high resolution polarimetric observations obtained by GST/NIRIS.
\end{abstract}

\keywords{Sun: magnetic fields $-$ Methods: data analysis $-$ Techniques: spectroscopic}

\section{Introduction} 
\label{sec:intro}

Stokes inversion has been an important yet challenging task in solar physics for decades 
\citep{Auer1977, Del1996, Asensio2015}.
Its purpose is to infer physical parameters such as the total magnetic field strength, 
inclination and azimuth angles, Doppler shift of the line center and so on from spectropolarimetric data. 
In general, such an inversion task is accomplished by attempting to find an appropriate forward model 
that best describes the relationship between the spectral shapes of the four Stokes components and the physical parameters, 
which is essentially a nonlinear nonconvex inverse problem. 
In the past, several inversion models have been developed. 
Based on the Levenberg-Marquardt algorithm \citep{Landolfi1984, Skumanich1987, Press1989}, 
a simplified model named the Milne-Eddington (ME) method \citep{Auer1977, Degl'Innocenti2004} 
provides an analytical solution for fast evaluation of the required derivatives in the algorithm.
Later, a more sophisticated method was introduced by \citet{Ruiz1992} based on response functions, 
which is able to retrieve height dependent information. 
This method has several different implementations including SPINOR \citep{Frutiger2000}, 
Helix+ \citep{Lagg2004} and VFISV \citep{Borrero2010}. 

In recent years, with rapid developments of advanced instruments and high-performance computers, 
powerful telescopes, such as the Daniel K. Inouye Solar Telescope \citep[\textit {DKIST};][]{McMullin2012}, 
European Solar Telescope \citep[\textit {EST};][]{Collados2008} 
and Goode Solar Telescope \citep[\textit {GST};][]{2012ASPC..463..357G} 
at the Big Bear Solar Observatory (BBSO), 
can produce data in unprecedented spatial and spectral resolution with high cadence.
In order to process these data in a time that is practical on a human timescale, 
more efficient and stable automated methods are in demand.
Many researchers have demonstrated that it is effective and efficient to perform Stokes inversion based on machine learning. 
For example, \citet{Socas-Navarro2001}, \citet{Ruiz2013}, and \citet{Quintero2015} developed methods 
for transforming Stokes profiles to a low-dimensional space using principal component analysis, 
which reduces the computational load and makes subsequent inversions faster. 
\citet{Carroll2001}, \citet{Socas-Navarro2003, Socas-Navarro2005}, and \citet{Carroll2008} 
employed multilayer perceptrons (MLP) for Stokes inversion,
demonstrating the speed, noise tolerance and stability of the MLP. 
\citet{Rees2004} and \citet{Teng2015}
used multiple support vector regression (MSVR) for real-time Stokes inversion. 
More recently, \citet{Asensio2019} performed Stokes inversion based on convolutional neural networks \citep[CNNs;][]{LeCun2015}
and applied their techniques to synthetic Stokes profiles obtained from snapshots of three-dimensional magneto-hydrodynamic 
numerical simulations of different structures of the solar atmosphere.

In this paper, we present a new machine learning method, also based on CNNs, 
for Stokes inversion on the Near InfraRed Imaging Spectropolarimeter (NIRIS) data \citep{2012ASPC..463..291C}. 
Our CNN method differs from that of \citet{Asensio2019} in two ways.
First, \citet{Asensio2019} used Stokes spectra synthesized in 3D MHD simulations of the solar atmosphere
and employed the CNNs to exploit all the spatial information encoded in a training dataset.
In contrast, our method performs pixel-by-pixel inversions, exploiting the spatial information of the Stokes profiles in a pixel. 
Second, in the synthetic data used by \citet{Asensio2019}, each Stokes component has 112 spectral points.
In contrast, in our NIRIS data, each Stokes component has 60 spectral points. 
Due to the different input sizes, the architecture of our CNN is different from those in \citet{Asensio2019}.

The rest of this paper is organized as follows. 
Section 2 describes the NIRIS data used in this study and our data collection scheme. 
Section 3 details our proposed CNN architecture and algorithm. 
Section 4 reports experimental results. 
Section 5 concludes the paper.

\section{Data}
\label{sec:data}
\begin{figure}
	\centering
	\includegraphics[width=7in]{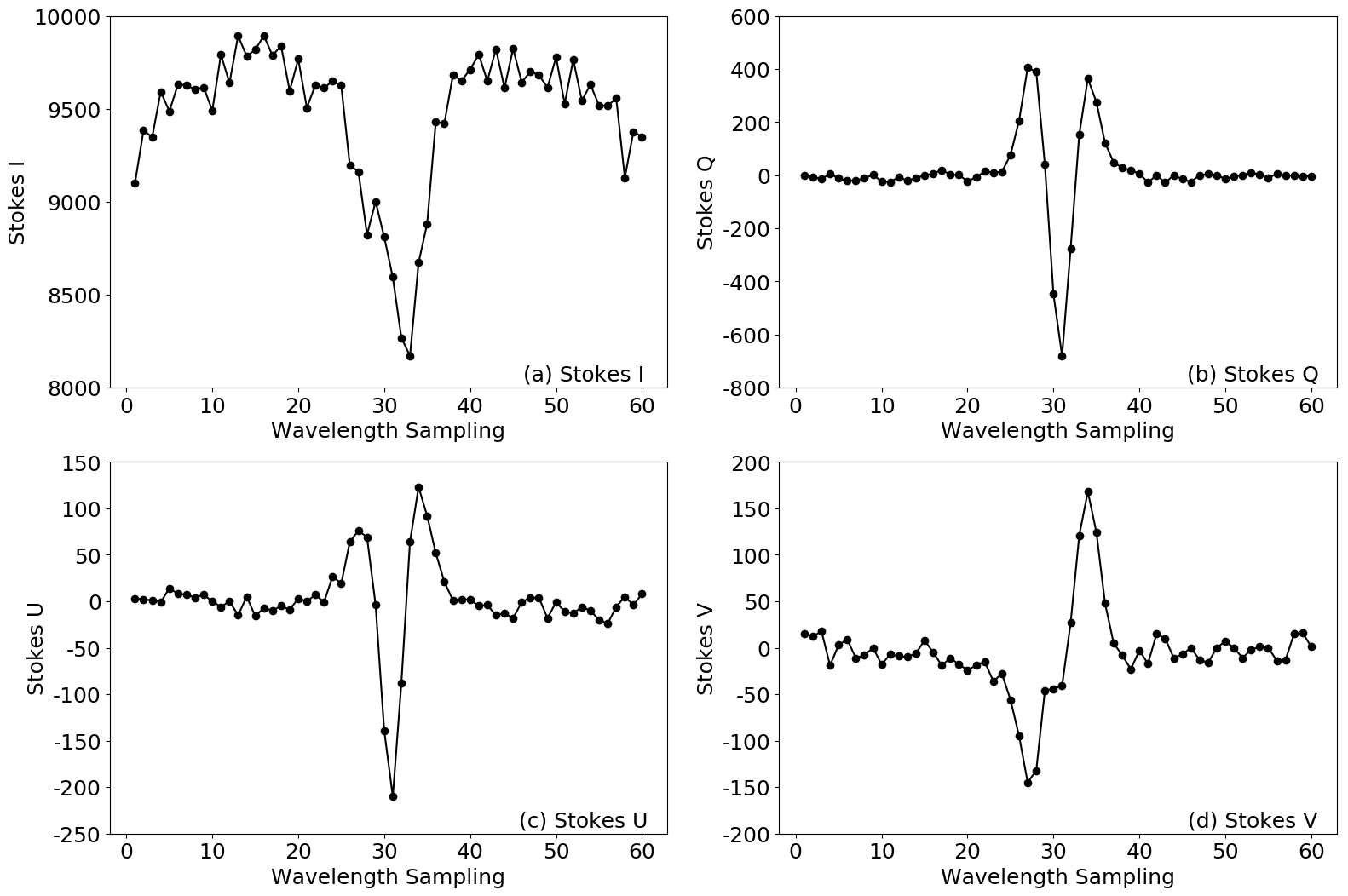}
	\caption{Stokes profiles of a pixel with a 857 Gauss magnetic field strength, 
          98 degree inclination angle, and 8 degree azimuth angle calculated by the ME method. 
         Each Stokes component has 60 wavelength sampling points.}
	\label{fig:Stokes}
\end{figure}

The GST/NIRIS is the second generation of the InfraRed Imaging Magnetograph \citep[IRIM;][]{Cao2006}, 
offering unprecedented high resolution vector magnetograms of the solar atmosphere 
from the deepest photosphere through the base of the corona. 
Its dual Fabry-P$\acute{\text{e}}$rot etalons provide 
an 85 arcsec field-of-view (FOV) 
with a cadence of 1 sec for spectroscopic scan and 10 sec for full Stokes measurements. 
The system utilizes half the chip to capture two simultaneous polarization states side-by-side, 
and provides an image scale of 
0$\farcs$083/pixel. 
It produces full spectroscopic measurements I, Q, U, V (Stokes profiles) 
at a spectral resolution of 0.01 nm in Fe I 1564.8 nm band, 
with a typical range of $-0.25$ to $+0.25$ nm from the line center \citep{Wang2015, Xu2016,  Wang2017, Liu2018, Xu2018}. 
Figure \ref{fig:Stokes} illustrates the Stokes I, Q, U, V components of a pixel
with a 857 Gauss magnetic field strength, 98 degree inclination angle, 
and 8 degree azimuth angle calculated by the
 Milne-Eddington (ME) method \citep{Auer1977, Degl'Innocenti2004}.
Each Stokes component contains 60 wavelength sampling points. 

We consider three active regions (ARs), namely
AR 12371,
AR 12665 and
AR 12673,
 in four different days.
For the AR 12371, we consider ten 990$\times$950 images collected at ten different time points on 2015 June 22; 
we randomly select one million pixels (data samples) from these ten images to form the training set.
Then, again for the AR 12371, we consider ten 720$\times$720 images collected at ten different time points on 2015 June 25;
we use the image collected at 20:00:00 UT on 2015 June 25 as the first test set.
Next, we consider ten 720$\times$720 images from the AR 12665 collected at ten different time points on 2017 July 13;
we use the image collected at 18:35:00 UT on 2017 July 13 as the second test set.
Finally, we consider one 720$\times$720 image from the AR 12673 collected at 19:18:00 UT on 2017 September 6, and
use this image as the third test set.
Each test set (image) has 518400 pixels corresponding to 518400 data samples.
The training set and each of the test sets are disjoint. 
The first test set is of the same active region and within $\sim$3 days of the training set, while 
the second test set and third test set are of different active regions, just over 2 years later. 
We want to see how well the trained CNN model works on these different test sets.

Each data sample (pixel) is comprised of Stokes I, Q, U, V profiles taken at 60 spectral points.
In addition, each data sample has a label, which is the vector magnetic field,
including the total magnetic field strength, inclination angle and azimuth angle,
calculated by the ME method. 
During training, the labels of the data samples in the training set are used to train and optimize our CNN model. 
Because the labels of the training data are created by the physics-based ME method,
our CNN model can be considered as a physics-assisted deep learning-based method.

During testing, we use the trained CNN model to predict or infer the label 
of a test data sample
from the Stokes Q, U, V profiles, calibrated by the Stokes I component \citep{Unno1956}, 
of the test data sample.
We then compare the labels (i.e., vector magnetic fields)
 inferred by our CNN model with those calculated by the ME method
for the test data samples under consideration.
Because the Stokes profiles and labels have different units and scales, 
we normalize them as follows. 
For the Stokes profiles, we normalize them by dividing them by 1000. 
For the labels, we normalize the total magnetic field strength by dividing it by 5000, and
normalize the inclination angle and azimuth angle by dividing them by $\pi$ respectively. 
The two numbers, 1000 and 5000, are used here because most of the Stokes measurements
have values  between $-1000$ and $+1000$, and their total magnetic field strengths range from $-5000$  Gauss to $+5000$  Gauss.

After obtaining the estimated vector magnetic field, which is inferred by our trained model, of a test data sample (pixel),
we can derive the three Cartesian components of the magnetic field, namely $B_x$, $B_y$ and $B_z$, of the pixel as follows:
\begin{equation} \label{eq:1}
\left\{ \begin{array}{ll}
&B_x=B_{total} \times \text{sin}\phi \times \text{cos}\theta \\
&B_y=B_{total} \times \text{sin}\phi \times \text{sin}\theta \\
&B_z=B_{total} \times \text{cos}\phi
\end{array} \\ \right.
\end{equation}
where $B_{total}$ denotes the total magnetic field strength, 
$\phi$ is the inclination angle, 
and $\theta$ is the azimuth angle. 

\section{Methodology}
\begin{figure}
	\centering
	\includegraphics[width=5.5in]{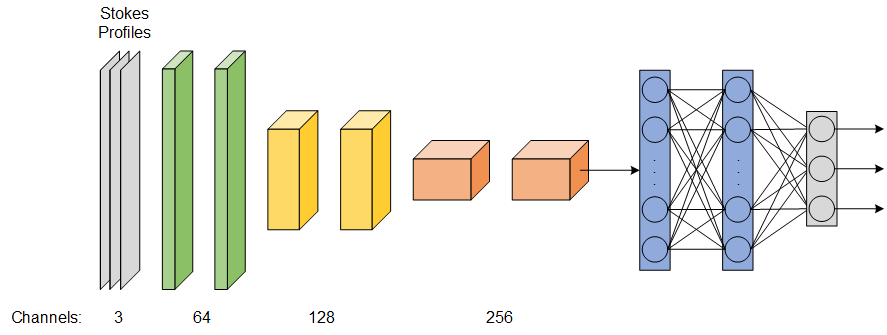}
	\caption{Architecture of our convolutional neural network (CNN). 
          This network is comprised of an input layer, three convolutional blocks, two fully connected layers and an output layer. 
         The input of the CNN is a three-channels sequence of Stokes Q, U, V  components each having 60 wavelength sampling points. 
        The intermediate outputs of the three convolutional blocks have 64, 128 and 256 channels respectively. 
       There are 1024 neurons activated by ReLU in both of the two fully connected layers.
        The output layer has three neurons activated by the Tanh function,
        where each neuron produces a value in the range $(-1, 1)$
        representing the total magnetic field strength, inclination angle and azimuth angle, respectively.}
	\label{fig:architecture}
\end{figure}

We use a convolutional neural network (CNN) to  infer vector magnetic fields from Stokes profiles of GST/NIRIS.
Our CNN model helps in denoising inversions by exploiting the spatial information of the Stokes profiles.
 Figure \ref{fig:architecture} presents the architecture of our network.
 It contains an input layer, three convolutional blocks, two fully connected layers and an output layer. 
The input layer receives a sequence of 
Stokes Q, U, V  components, each having 60 wavelength sampling points, 
with 3 channels.
Each channel corresponds to a Stokes component respectively.

After the input layer, there are three convolutional blocks with the following structures. 
The first convolutional block consists of two convolutional layers, 
which take, as input, the output from the previous layer 
and filter it with 64 kernels of sizes 3$\times$1$\times$3 and 3$\times$1$\times$64 respectively, 
and a max-pooling layer with a pooling factor of 2. 
The second convolutional block consists of two convolutional layers 
with filters of 128 kernels of sizes 3$\times$1$\times$64 and 3$\times$1$\times$128 respectively, 
and a max-pooling layer with a pooling factor of 2. 
The third convolutional block consists of two convolutional layers 
with filters of 256 kernels of sizes 3$\times$1$\times$128 and 3$\times$1$\times$256 respectively. 
The third convolutional block does not contain a max-pooling layer.

The activation functions used in both the convolutional layers and fully connected layers
are rectified linear units \citep[ReLU;][]{DBLP:books/daglib/0040158}, defined as:
\begin{equation}
\text{ReLU}(x)=\text{max}(0, x)=\left\{ \begin{array}{ll}
					x \qquad \qquad &\text{if} \  x \geq 0\\
 					0 \qquad \qquad &\text{if} \ x < 0
					\end{array} \\ \right.
\end{equation}
The output of the three convolutional blocks is flattened into a sequence,
which is then sent to the two fully connected layers each having 1024 neurons activated by ReLU. 
Finally, there is an output layer with 3 neurons activated by the hyperbolic tangent function \citep[Tanh;][]{DBLP:books/daglib/0040158},
defined as: 
\begin{equation}
\text{Tanh}(x)=\left(\frac{e^x-e^{-x}}{e^x+e^{-x}}\right),
\end{equation}
where each neuron outputs a value that lies in the range $(-1, 1)$
representing the total magnetic field strength, inclination angle and azimuth angle, respectively.
The training of the CNN model is done by optimizing L1 loss defined as follows
\citep{DBLP:books/daglib/0040158}:
\begin{equation} \label{eq:lossFunc}
\text{L1 loss} =\frac{1}{N}\sum_{i=1}^N (|y^{\text{tot}}_{i} - \hat{y}^{\text{tot}}_{i}| + 
|y^{\text{inc}}_{i} - \hat{y}^{\text{inc}}_{i}| + |y^{\text{azi}}_{i} - \hat{y}^{\text{azi}}_{i}|),
\end{equation}
where $N = 1,000,000$ is the total number of pixels in the training set, and
$y^{\text{tot}}_{i}$, $y^{\text{inc}}_{i}$, $y^{\text{azi}}_{i}$  ($\hat{y}^{\text{tot}}_{i}$, $\hat{y}^{\text{inc}}_{i}$, $\hat{y}^{\text{azi}}_{i}$ respectively) 
denotes the total magnetic field strength, inclination angle and azimuth angle of the $i$th pixel calculated by the ME method
(inferred by our CNN method, respectively). 
L1 loss is chosen here because it is efficient and produces good results as shown in Section \ref{results}.

Our CNN model is implemented in Python, TensorFlow and Keras. 
A mini-batch strategy \citep{LeCun2015, DBLP:books/daglib/0040158} is used 
to achieve faster convergence during backpropagation. 
The optimizer used is Adam \citep{LeCun2015, DBLP:books/daglib/0040158}, 
which is a stochastic gradient descent method.
The initial learning rate is set to 0.001 with a learning rate decay of 0.01 over each epoch, 
$\beta_1$ is set to 0.9, and $\beta_2$ is set to 0.999. 
The batch size is set to 256 and the number of epochs is set to 50.

During testing, to infer the physical parameters of each pixel in a test image, 
we take the Stokes Q, U, V profiles of the pixel and feed them to the trained CNN model. 
The CNN model will output a three-dimensional vector with normalized values in the range $(-1, 1)$
representing the total magnetic field strength ($B_{total}$), inclination angle ($\phi$)  
and azimuth angle ($\theta$) respectively.
By de-normalization of the values, we can obtain the inferred or estimated 
$B_{total}$, $\phi$ and $\theta$
of the pixel.
 Furthermore, based on the estimated $B_{total}$, $\phi$ and $\theta$, 
we can derive the three Cartesian components of the magnetic field, 
namely $B_x$, $B_y$ and $B_z$, of the pixel using Equation \eqref{eq:1}.

\section{Results}
\label{results}
\subsection{Performance Metrics}
We conducted a series of experiments to evaluate the performance of the proposed CNN model and 
compare it with related methods based on four performance metrics: 
mean absolute error \citep[MAE;][]{SS-1990}, 
percent agreement \citep[PA;][]{McHugh2012},
R-squared \citep{SS-1990} and
Pearson product-moment correlation coefficient \citep[PPMCC;][]{Galton1886, Pearson1895}.
We considered six quantities:
total magnetic field strength ($B_{total}$),
 inclination angle ($\phi$),
 azimuth angle ($\theta$),
 $B_x$, $B_y$ and $B_z$.
For each quantity, we compared its ME-calculated values with our CNN-inferred values and computed the four performance metrics.

The first performance metric is defined as \citep{SS-1990}:
\begin{equation} \label{eq:MAE}
\text{MAE} =\frac{1}{N}\sum_{i=1}^N |y_{i} - \hat{y}_{i}|,
\end{equation}
where $N$ is the total number of data samples (pixels) in a test image, and
$y_{i}$  ($\hat{y}_{i}$, respectively) 
denotes the ME-calculated
(CNN-inferred, respectively) value for the $i$th pixel in the test image.
This metric is used to quantitatively assess the dissimilarity (distance) between the ME-calculated values and CNN-inferred values 
in the test image.
The smaller the MAE is, the better performance a method has.

The second performance metric is defined as \citep{McHugh2012}:
\begin{equation}
\text{PA}=\frac{M}{N} \times 100\%,
\end{equation}
where $M$ denotes the total number of agreement pixels in the test image.
We say the $i$th pixel in the test image is an agreement pixel if 
$|y_{i}-\hat{y}_{i}|$ is smaller than a user-specified threshold. 
(The default thresholds are set to 200 Gauss for $B_{total}$, $B_x$, $B_y$, $B_z$ respectively
and 10 degree for $\phi$, $\theta$ respectively.)
This metric is used to quantitatively assess the similarity between the
ME-calculated values and CNN-inferred values 
in the test image.
The larger the PA is, the better performance a method has.

The third performance metric is defined as  \citep{SS-1990}:
\begin{equation}
\text{R-squared} =1 - \frac{\sum_{i=1}^{N}(y_i - \hat{y}_i)^2}{\sum_{i=1}^{N}(y_i-\overline{y})^2},
\end{equation}
where $\overline{y}=\frac{1}{N}\sum_{i=1}^{N}y_i$ denotes the mean of the ME-calculated values
for all the pixels in the test image.
The R-squared value, ranging  from $-\infty$ to 1,  is used to measure the strength of the relationship 
between the ME-calculated values and CNN-inferred values 
in the test image.
The larger (i.e., the closer to 1) the R-squared value is, 
 the stronger relationship 
between the ME-calculated values and CNN-inferred values we have.
 
 The fourth performance metric is defined as \citep{Galton1886, Pearson1895}:
 \begin{equation}
 \text{PPMCC} =\frac{\text{E}[(X-\mu_X)(Y-\mu_Y)]}{\sigma_X \sigma_Y},
 \end{equation}
 where $X$ and $Y$ represent the ME-calculated values and CNN-inferred values respectively, 
 $\mu_X$ and $\mu_Y$ are the mean of $X$ and $Y$ respectively,
 $\sigma_X$ and $\sigma_Y$ are the standard deviation of $X$ and $Y$ respectively,
 and $E(\cdot)$ is the expectation.
 The value of PPMCC ranges from $-1$ to 1.
 A value of 1 means that a linear equation describes the relationship between $X$ and $Y$ perfectly
 where all data points lying on a line for which $Y$ increases as $X$ increases.
 A value of $-1$ means that all data points lie on a line for which $Y$ decreases as $X$ increases.
 A value of 0 means that there is no linear correlation between the variables $X$ and $Y$.
We will mainly use PPMCC in our experimental study because 
it measures the linear correlation between the ME-calculated values and  CNN-inferred values, 
quantifying how well the CNN-inferred values agree with the ME-calculated values in the test image \citep{Galton1886, Pearson1895, SS-1990}. 
The larger (i.e., the closer to 1) the PPMCC is, the better performance a method has.
Notice that PA, R-squared and PPMCC do not have units while
MAE has units: ``Gauss" for $B_{total}$, $B_x$, $B_y$, $B_z$ respectively and
``degree" for $\phi$ (inclination angle), $\theta$ (azimuth angle) respectively. 

\subsection{Results of Using AR 12371 on 2015 June 22 as Training Data}
In this experiment, we used the one million data samples (pixels) from AR 12371 
collected on 2015 June 22 as the training data to train our CNN model.
We then used the trained CNN model to infer vector magnetic fields from the Stokes Q, U, V profiles of the pixels in the 
three test sets (images) described in 
Section \ref{sec:data}.\footnote{The source code and datasets used in the experiment can be downloaded from \url{https://web.njit.edu/~wangj/CNNStokesInversion/}.}
For comparison purposes, we also used the 
Milne-Eddington (ME) method \citep{Auer1977, Degl'Innocenti2004}
to derive the vector magnetic fields of the pixels in the three test images.

\begin{figure}
	\centering
	\gridline{\fig{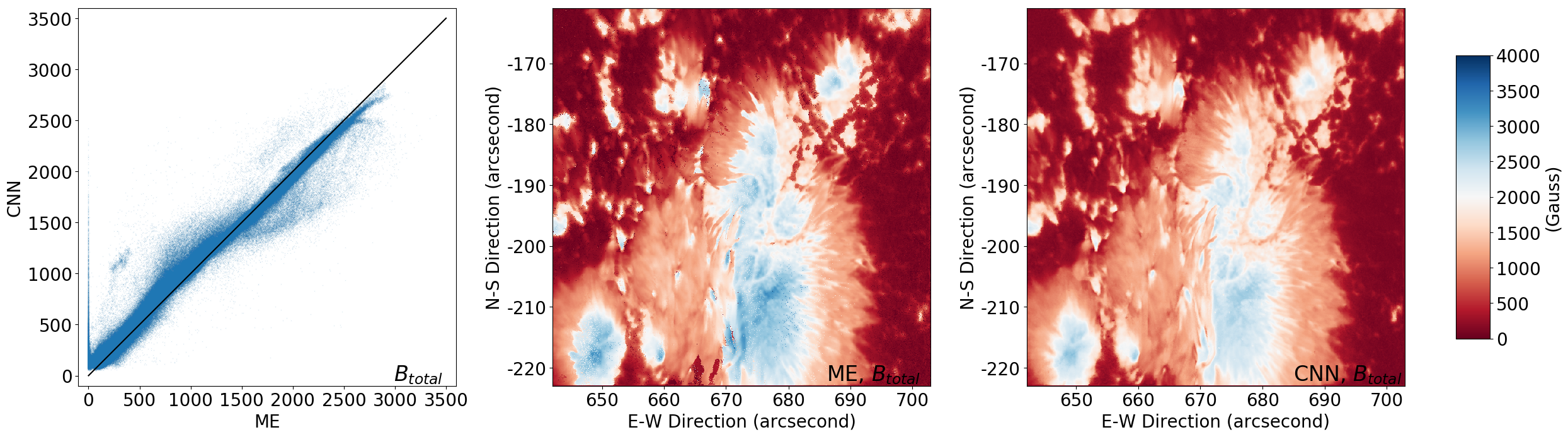}{\textwidth}{}}
	\gridline{\fig{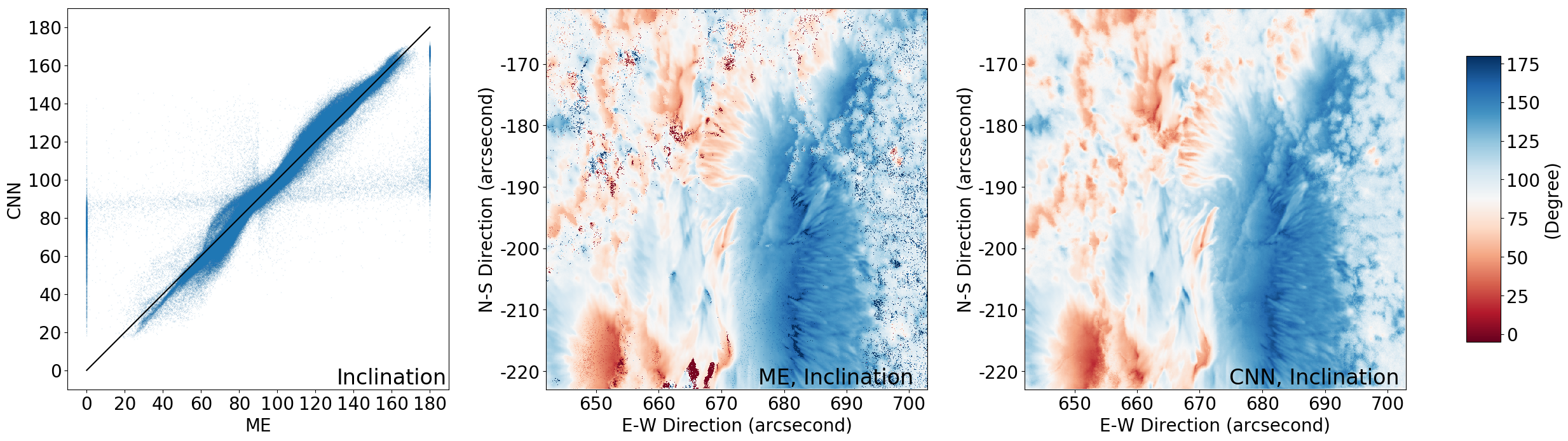}{\textwidth}{}}
	\gridline{\fig{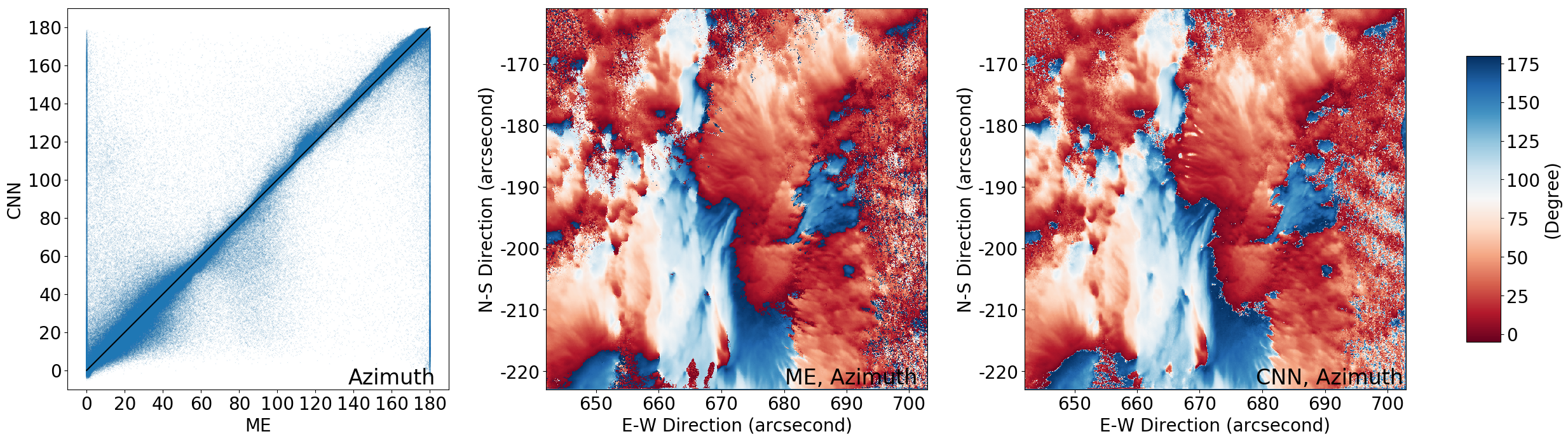}{\textwidth}{}}
	\caption{Comparison between the ME and CNN methods for deriving $B_{total}$, $\phi$ (inclination angle) and $\theta$ (azimuth angle) 
		based on the test image from AR 12371 collected on 2015 June 25 20:00:00 UT 
		where training data were taken from the same AR 12371 on 2015 June 22. 
		Displayed from top to bottom are the results for $B_{total}$, $\phi$ (inclination angle) and $\theta$ (azimuth angle) respectively. 
		The first column shows scatter plots where the X-axis and Y-axis represent the values obtained by the ME and CNN methods respectively. 
		The black diagonal line in each scatter plot corresponds to pixels whose ME-calculated values are identical to CNN-inferred values. 
		The second column shows magnetic maps derived by the ME method. 
		The third column shows magnetic maps inferred by our CNN method.}
	\label{fig:20150625_200000-1} 
\end{figure}

\begin{figure}
	\gridline{\fig{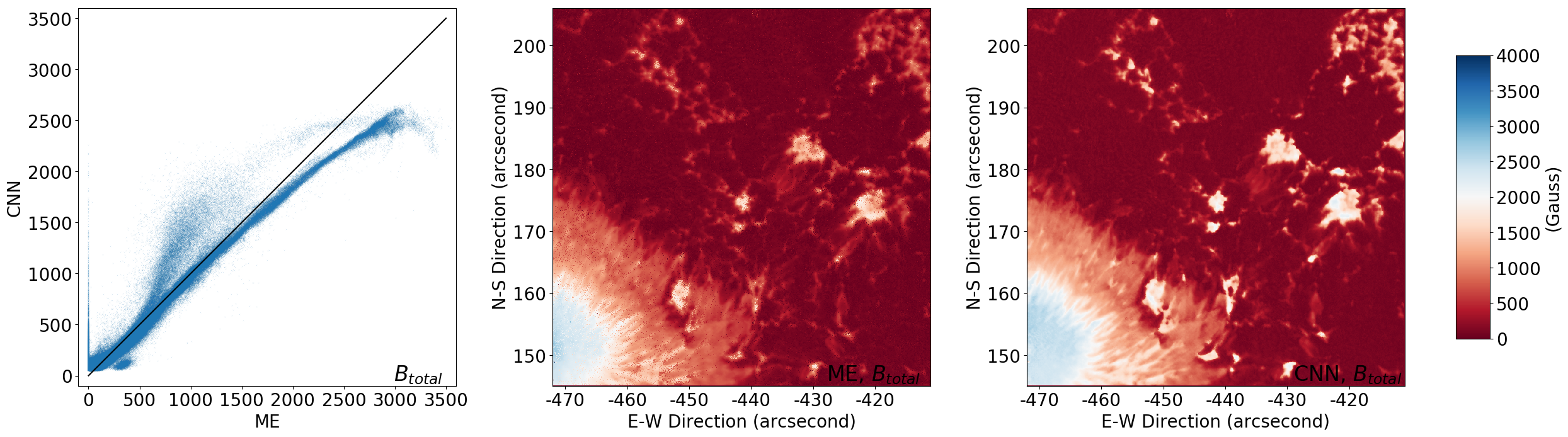}{\textwidth}{}}
	\gridline{\fig{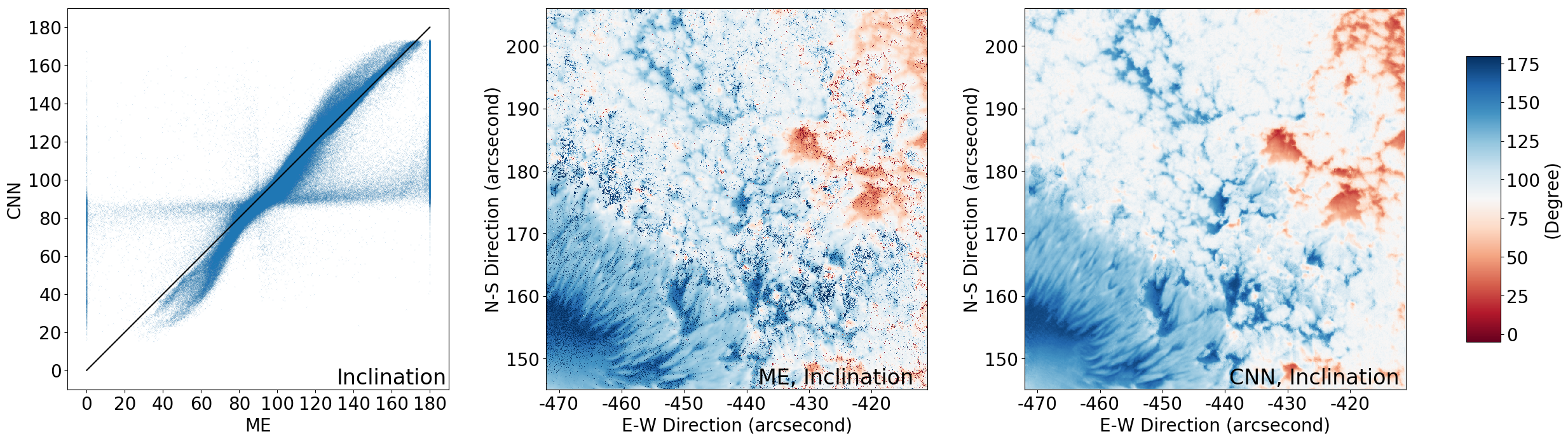}{\textwidth}{}}
	\gridline{\fig{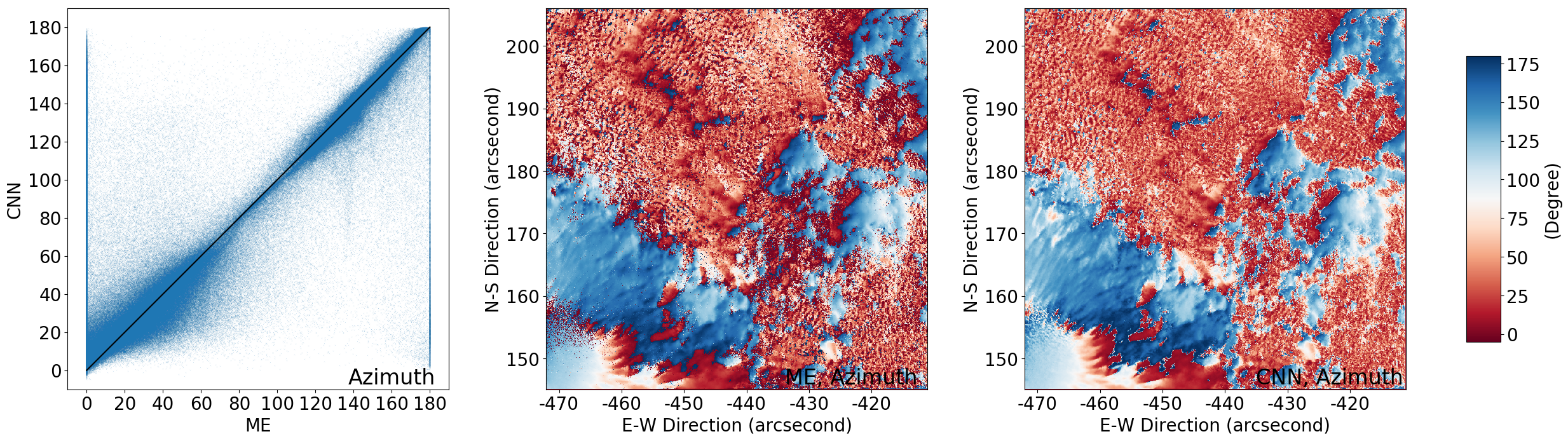}{\textwidth}{}}
	\caption{Comparison between the ME and CNN methods for deriving $B_{total}$, $\phi$ (inclination angle) and $\theta$ (azimuth angle) 
		based on the test image from AR 12665 collected on 2017 July 13 18:35:00 UT 
		where training data were taken from AR 12371 on 2015 June 22. 
		Displayed from top to bottom are the results for $B_{total}$, $\phi$ (inclination angle) and $\theta$ (azimuth angle) respectively. 
		The first column shows scatter plots where the X-axis and Y-axis represent the values obtained by the ME and CNN methods respectively. 
		The black diagonal line in each scatter plot corresponds to pixels whose ME-calculated values are identical to CNN-inferred values. 
		The second column shows magnetic maps derived by the ME method. 
		The third column shows magnetic maps inferred by our CNN method. }
	\label{fig:20170713_183500-1} 
\end{figure}

\begin{figure}
	\gridline{\fig{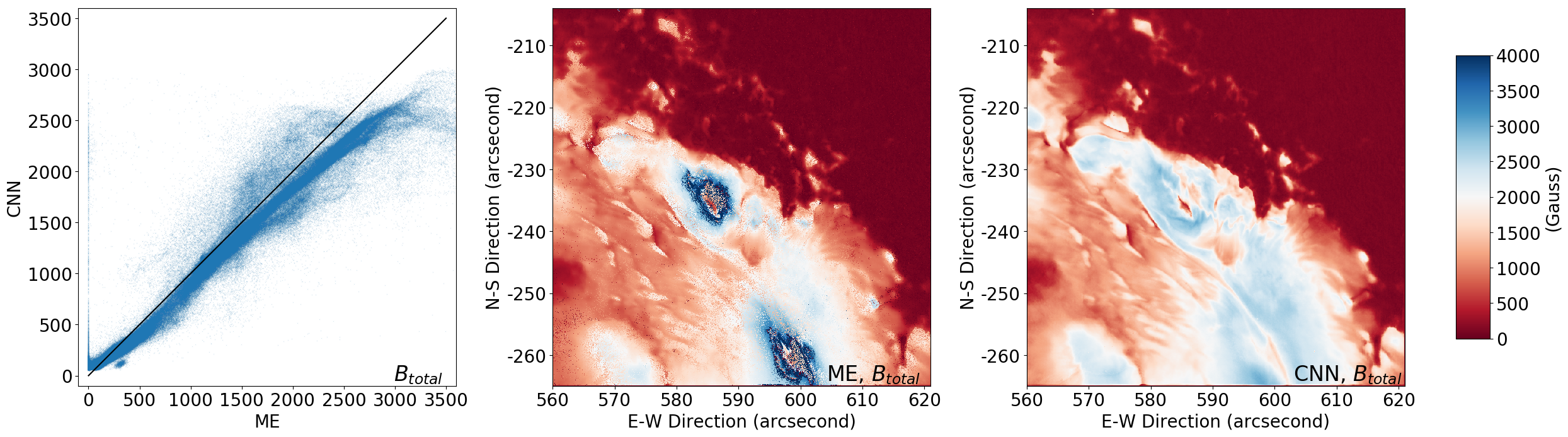}{\textwidth}{}}
	\gridline{\fig{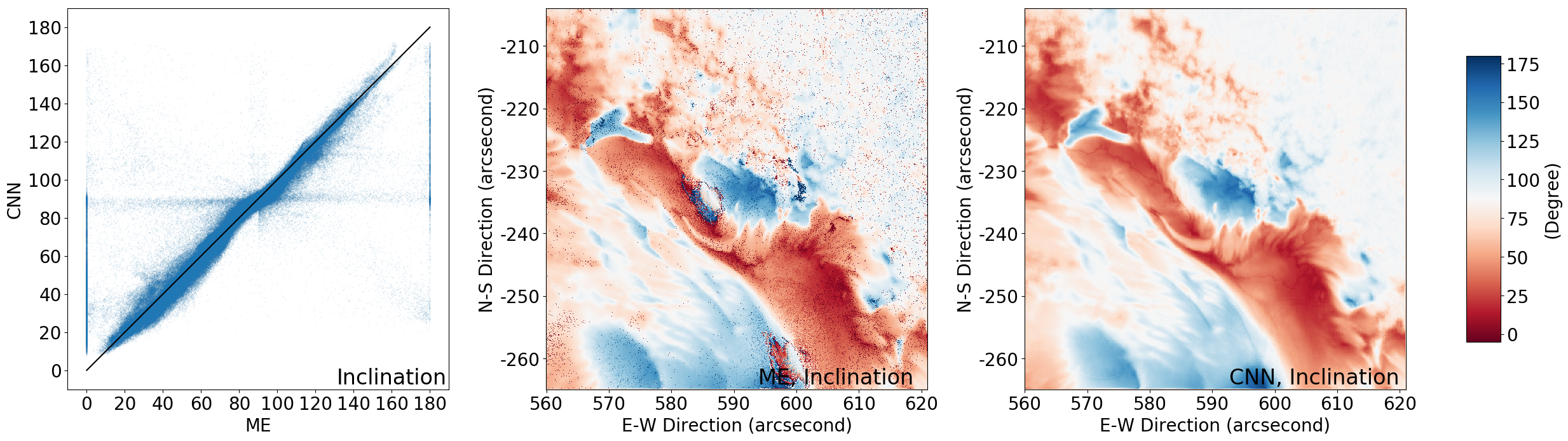}{\textwidth}{}}
	\gridline{\fig{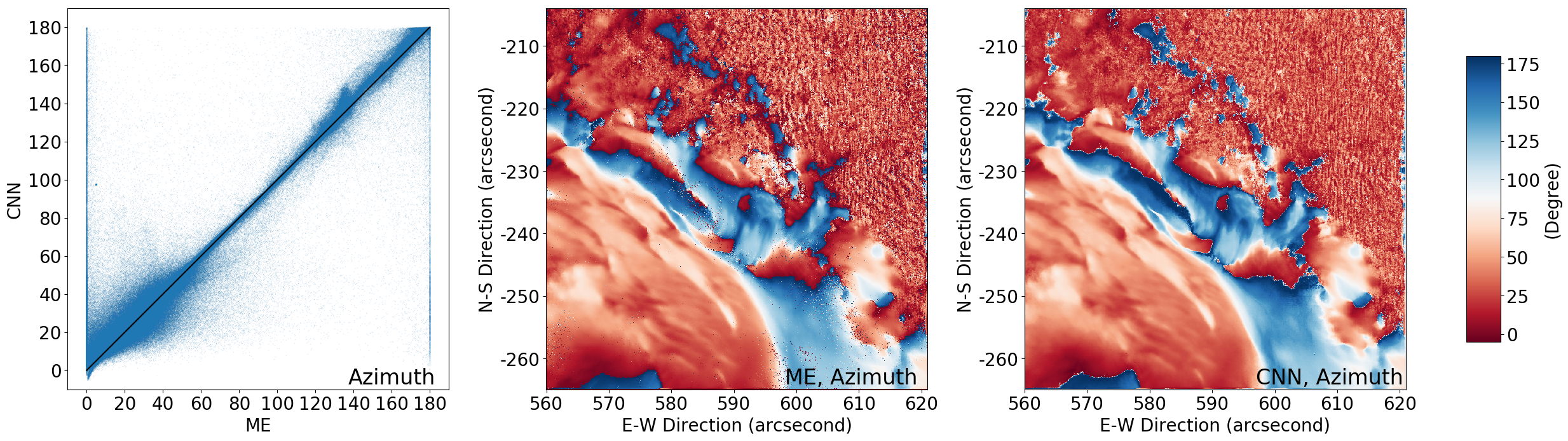}{\textwidth}{}}
	\caption{Comparison between the ME and CNN methods for deriving $B_{total}$, $\phi$ (inclination angle) and $\theta$ (azimuth angle) based on the test image from AR 12673 collected on 2017 September 6 19:18:00 UT where training data were taken from AR 12371 on 2015 June 22. 
		Displayed from top to bottom are the results for $B_{total}$, $\phi$ (inclination angle) and $\theta$ (azimuth angle) respectively. 
		The first column shows scatter plots where the X-axis and Y-axis represent the values obtained by the ME and CNN methods respectively. 
		The black diagonal line in each scatter plot corresponds to pixels whose ME-calculated values are identical to CNN-inferred values. 
		The second column shows magnetic maps derived by the ME method. 
		The third column shows magnetic maps inferred by our CNN method. }
	\label{fig:20170906_191800-1} 
\end{figure}

Figure \ref{fig:20150625_200000-1} (Figure \ref{fig:20170713_183500-1}, Figure \ref{fig:20170906_191800-1} respectively)
presents results for the three obtained quantities $B_{total}$, $\phi$ (inclination angle) and $\theta$ (azimuth angle),
displayed from top to bottom in the figure,
of the test image with 720$\times$720 pixels from AR 12371 (AR 12665, AR 12673 respectively)
collected on 2015 June 25 20:00:00 UT (2017 July 13 18:35:00 UT, 2017 September 6 19:18:00 UT respectively).
In all the figures, the first column shows scatter plots for each obtained quantity.
The X-axis and Y-axis in each scatter plot represent the values
obtained by the ME method and CNN method respectively. 
The black diagonal line in each scatter plot 
corresponds to pixels whose ME-calculated values are identical to CNN-inferred values.
The second columns in these figures
show magnetic maps with 720$\times$720 pixels derived by the ME method.
The third columns in the figures show magnetic maps with 720$\times$720 pixels inferred by our CNN method.

\begin{figure}
	\includegraphics[width=\textwidth]{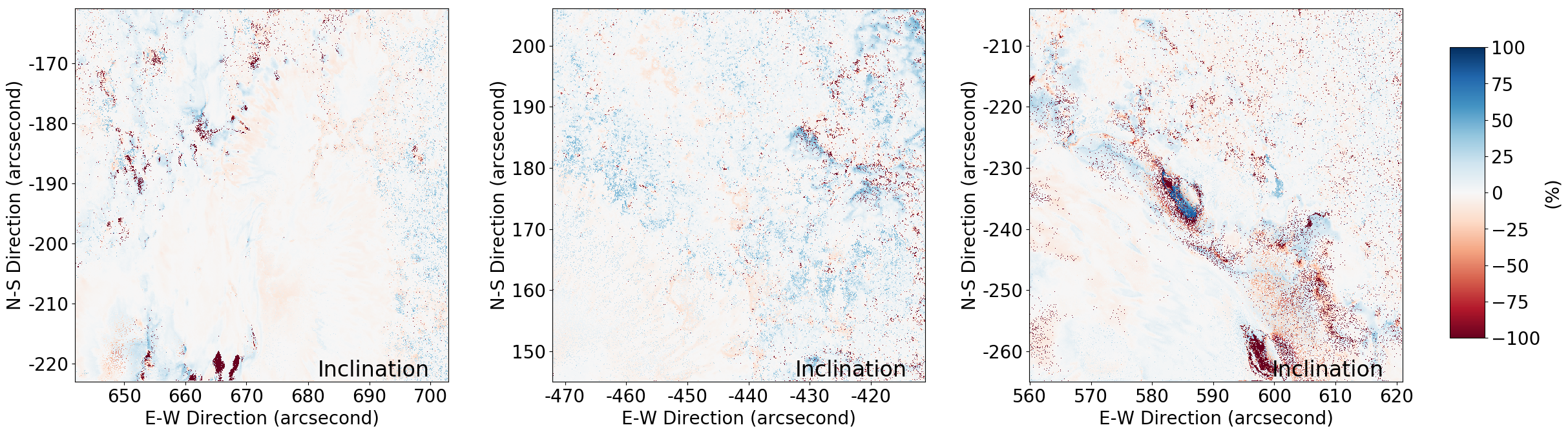}
	\caption{Percentage difference images for the $\phi$ (inclination angle) maps. 
				The first column shows the percentage difference image based on the test image from AR 12371 collected on 2015 June 25 20:00:00 UT.
				The second column shows the percentage difference image based on the test image from AR 12665 collected on 2017 July 13 18:35:00 UT.
				The third column shows the percentage difference image based on the test image from AR 12673 collected on 2017 September 6 19:18:00 UT.
				These percentage difference images highlight the locations of the differences between the CNN-inferred $\phi$ values and ME-calculated $\phi$ values
				in the three test images.}
	\label{fig:diff-image} 
\end{figure}

\textbf{Summary of the results.}
The scatter plots in the figures show that
the Stokes inversion results obtained by our CNN method and the ME method are highly correlated. 
From the top-left panels in Figures \ref{fig:20150625_200000-1}, \ref{fig:20170713_183500-1} and \ref{fig:20170906_191800-1},
we see that the CNN-inferred $B_{total}$ values are closer to the ME-calculated $B_{total}$ values in the low-field end
and are farther from the ME-calculated $B_{total}$ values in the high-field end.
The figures also show that the CNN method produces smoother and cleaner magnetic maps than the ME method.
There are salt-pepper noise pixels in the magnetic maps produced by the ME method.
To help locate the noise pixels, we use percentage difference images in which the value of the $i$th pixel is
equal to $(y_{i} - \hat{y}_{i})$/$y_{i}$ $\times$ 100\% where 
$y_{i}$  ($\hat{y}_{i}$, respectively) 
denotes the ME-calculated
(CNN-inferred, respectively) value for the $i$th pixel.
For example, Figure \ref{fig:diff-image} shows the percentage difference images for the $\phi$ (inclination angle) maps 
in Figures \ref{fig:20150625_200000-1}, \ref{fig:20170713_183500-1} and \ref{fig:20170906_191800-1}.
The percentage difference images highlight the locations of the differences between the CNN-inferred $\phi$ values and ME-calculated $\phi$ values
in the test images.
Figure \ref{fig:20150625_200000-2} (Figure \ref{fig:20170713_183500-2}, Figure \ref{fig:20170906_191800-2} respectively) in the Appendix
presents results for the 
quantities $B_x$, $B_y$ and $B_z$, 
displayed from top to bottom in the figure, 
of the test image with 720$\times$720 pixels from AR 12371 (AR 12665, AR 12673 respectively)
collected on 2015 June 25 20:00:00 UT (2017 July 13 18:35:00 UT, 2017 September 6 19:18:00 UT respectively).

To quantitatively assess the number of noise pixels in the magnetic maps derived by the ME and CNN methods, 
we adopt a threshold-based algorithm, which works as follows.
We define $P$ to be a noise pixel (outlier) with respect to a user-specified threshold 
if among $P$'s eight neighboring pixels,
there are more than four neighboring pixels
satisfying the following condition: the difference between the value of a neighboring pixel and
the value of $P$ is greater than or equal to the threshold.
The default thresholds are set to 500 Gauss for 
$B_{total}$, $B_x$, $B_y$, $B_z$ respectively
and 20 degree for $\phi$ (inclination angle), $\theta$ (azimuth angle) respectively.
We define the outlier-difference to be
the number of outliers produced by the ME method minus the number of outliers produced by our CNN method.
A positive outlier-difference means ME produces more outliers than CNN 
while a negative outlier-difference means CNN produces more outliers than ME.

Table \ref{tab:single-image-metrics} presents the performance metric values of the CNN method. 
The results in Table \ref{tab:single-image-metrics} are consistent with those 
in Figures \ref{fig:20150625_200000-1}-\ref{fig:20170906_191800-2}.
Specifically, the CNN-inferred results are highly correlated to the ME-calculated results with PPMCC values being close to 1.
Furthermore, CNN produces smoother magnetic maps with fewer outliers (noise pixels) than the ME method.
This happens because among the one million training data samples whose labels are calculated by the ME method,
there are relatively few outliers. 
The CNN method can learn latent patterns from the majority of the training data samples, which are clean.
As a consequence, we obtain a good CNN model capable of producing clean results.
Tables \ref{tab:multiple-images-metrics-20150625} and \ref{tab:multiple-images-metrics-20170713} in the Appendix
present the performance metric values for the test images from AR 12371 and AR 12665
collected at ten different time points on 2015 June 25 and 2017 July 13 respectively. 
The results in these tables are consistent with those in Table \ref{tab:single-image-metrics}.

\begin{table}[!htbp]
	\caption{Performance Metric Values of Our CNN Method Based on the Test Images from Three Active Regions}
	\label{tab:single-image-metrics}
	\centering
	\begin{tabular}{cl||c||c||c||c||c||c}
		\hline
		& & $B_{total}$  & \text{$B_x$}  & \text{$B_y$}  & \text{$B_z$} & $\phi$   & $\theta$   \\ \hline
		\multirow{5}{*}{2015-06-25 20:00:00 UT (AR 12371)}&MAE &  86.660 &  88.997 & 66.140 & 55.653 & 4.867 & 11.136  \\ 
		&PA &91.6\% &91.3\% &95.2\% &94.7\% &92.2\% &79.1\% \\
		&R-squared & 0.963 &  0.936& 0.901 & 0.976 & 0.838 & 0.720 \\  
		&PPMCC & 0.983 &  0.968 &0.951 &0.989 & 0.916 & 0.853 \\  
		&Outlier-difference &2959 &4380 &-770 &1050 &15108 &7219  \\ \hline
		\multirow{5}{*}{2017-07-13 18:35:00 UT (AR 12665)}&MAE & 73.684 &  71.555 & 51.170 & 49.023 & 7.573 & 17.437  \\ 
		&PA &91.5\% &93.3\% &96.4\% &92.6\% &84.8\% &60.6\% \\
		&R-squared & 0.950 &  0.841 & 0.851 & 0.941 & 0.663 & 0.665 \\ 
		&PPMCC & 0.976 &  0.918 & 0.926 & 0.971 &  0.827 & 0.821 \\  
		&Outlier-difference &3801 &7280 &3413 &2478 &35649 &28274  \\\hline
		\multirow{5}{*}{2017-09-06 19:18:00 UT (AR 12673)}&MAE &193.680 &146.100 &124.783 &136.892 &5.497 &9.009 \\ 
		&PA &75.0\% &80.1\% &86.2\% &87.2\% &91.3\% &79.1\% \\
		&R-squared &0.841 &0.884 &0.777 &0.736 &0.776 & 0.807\\ 
		&PPMCC &0.935 &0.943 &0.888 &0.859 &0.881 & 0.902 \\ 
		&Outlier-difference &19651 &22317 &16592 &12950 &21951 &14265  \\ \hline
	\end{tabular}
   \begin{threeparttable}
		\flushleft ~\textbf{Notes.} 
		\begin{tablenotes}
			\small
            \item[a] The performance metric values in the table are obtained by training the CNN model using one million pixels
            from AR 12371 collected on 2015 June 22 and then applying the trained model to
            the test image from AR 12371 collected on 2015 June 25 20:00:00 UT 
            (AR 12665 collected on 2017 July 13 18:35:00 UT, and AR 12673 collected on 2017 September 6 19:18:00 UT, respectively).
			\item[b] A positive outlier-difference means ME produces more outliers than CNN while a negative outlier-difference means CNN produces more outliers than ME.
		\end{tablenotes}
	\end{threeparttable}
\end{table}

\textbf{Comparison with related methods.} To further understand the behavior of our CNN method and compare it with related machine learning algorithms,
we conduct a cross-validation study as follows. 
We partition the training set of one million data samples from AR 12371 on 2015 June 22 into 10 equal-sized folds.
For every two training folds $i$ and $j$, $i$ $\neq$ $j$, fold $i$ and fold $j$ are disjoint.
The first test set contains the ten 720$\times$720 images, also from AR 12371, 
collected on 2015 June 25.
These test images are numbered from 1 to 10.
In run $i$, 1 $\leq$ $i$ $\leq$ 10, all training data samples except those in training fold $i$ 
are used to train a machine learning model, 
and the trained model is then used to make predictions on test image $i$.
We calculate the performance metrics MAE, PA, R-squared, PPMCC and outlier-difference based on the predictions made in run $i$. 
There are 10 runs. 
The means and standard deviations over the 10 runs are calculated and recorded. 
We also conduct the same cross-validation study for the second test set containing the ten 720$\times$720 images from AR 12665 
collected on 2017 July 13, and the third test set containing the 720$\times$720 image from AR 12673 collected on 2017 September 6. 
The third test set has only one image, and hence in each run, the same test image is used.

The related machine learning algorithms considered here include 
multiple support vector regression
 \citep[MSVR;][]{Rees2004, Teng2015} 
and multilayer perceptrons \citep[MLP;][]{Carroll2001, Socas-Navarro2003, Socas-Navarro2005, Carroll2008}. 
The MSVR method uses the radial basis function (RBF) kernel. 
The MLP model consists of an input layer, an output layer and two hidden layers both with 1024 neurons. 
Table \ref{tab:20150625-metrics} (Table \ref{tab:20170713-metrics}, Table \ref{tab:20170906-metrics} respectively) 
presents the mean MAE, PA, R-squared, PPMCC, outlier-difference and standard deviation
for each quantity $B_{total}$, $B_x$, $B_y$, $B_z$,
$\phi$ (inclination angle), $\theta$ (azimuth angle)
inferred by each of the three machine learning methods
MSVR, MLP and our CNN
for the first (second, third respectively) test set.
In the tables, PA, R-squared, PPMCC and outlier-difference do not have units while
MAE has units: ``Gauss" for $B_{total}$, $B_x$, $B_y$, $B_z$ respectively and
``degree" for $\phi$ (inclination angle), $\theta$ (azimuth angle) respectively.
It can be seen from
the tables that the CNN-inferred results are highly correlated to the ME-calculated results and are closer to the ME's results 
with PPMCC values being closer to 1 on average
than those from the other two machine learning methods.
In particular, based on the calculations 
on the six quantities $B_{total}$, $B_x$, $B_y$, $B_z$,
$\phi$ (inclination angle) and $\theta$ (azimuth angle) in Tables \ref{tab:20150625-metrics}-\ref{tab:20170906-metrics}, 
our CNN method outperforms the current best machine learning method (MLP) by 2.6\% on average in PPMCC.
However, there is no definite conclusion about outlier-differences among the three machine learning methods.

\begin{table}
	\centering
	\caption{Performance Metric Values of MSVR, MLP and Our CNN Method Based on the Test Set from AR 12371 Collected on 2015 June 25}
	\label{tab:20150625-metrics}
	\begin{tabular}{cl||c||c||c||c||c||c}
		\hline
		&     & $B_{total}$  & \text{$B_x$}  & \text{$B_y$}  & \text{$B_z$} & $\phi$   & $\theta$    \\ \hline
		\multirow{3}{*}{\parbox{1.5cm}{MAE}}
		& MSVR  &    437.02 (27.44) & 712.02 (19.03)  & 706.51 (12.24)  &  339.26 (17.86) & 23.02 (0.73)  &  84.43 (1.53)  \\
		& MLP  &   115.68 (5.15) &  109.44 (7.60) & 86.08 (4.80) & 80.62 (4.19) & 5.85 (0.31) & 12.29 (1.72)  \\
		& CNN  &  81.57 (3.66) & 76.56 (5.63) & 58.83 (2.86) & 52.18 (2.22) & 4.54 (0.23) & 9.34 (1.04)  \\ \hline
		\multirow{3}{*}{\parbox{1.5cm}{PA}}
		& MSVR  & 34.7\% (0.5\%) & 48.4\% (1.0\%) & 44.6\% (1.1\%) & 15.2\% (1.5\%) & 5.6\% (0.2\%) & 4.1\% (0.5\%) \\
		& MLP  & 86.2\% (0.7\%) & 88.4\% (0.8\%) & 91.5\% (0.5\%) & 89.5\% (0.7\%) & 89.4\% (1.0\%) & 76.7\% (1.0\%) \\
		& CNN  & 91.6\% (0.7\%) & 92.5\% (0.8\%) & 96.1\% (0.5\%) & 95.1\% (0.4\%) & 93.6\% (0.6\%) & 81.4\% (1.4\%) \\  \hline
		\multirow{3}{*}{\parbox{1.5cm}{R-squared}}
		& MSVR  & 0.45 (0.05)  & -0.92 (0.07)  & -5.34 (0.37)  & 0.28 (0.08) &  -0.09 (0.03)  &  -2.80 (0.16)  \\
		& MLP  &   0.92 (0.01) &  0.91 (0.01) & 0.85 (0.01) & 0.93 (0.01) &  0.80 (0.01) &  0.73 (0.04)  \\
		& CNN  &  0.97 (0.01) &  0.94 (0.01) & 0.93 (0.01) & 0.97 (0.01) &0.83 (0.01) &  0.76 (0.03)  \\  \hline
		\multirow{3}{*}{\parbox{1.5cm}{PPMCC}}
		& MSVR  & 0.82 (0.01)  & -0.09 (0.04)  & -0.10 (0.08)  &  0.89 (0.01)  & 0.85 (0.01)  & 0.48 (0.03)  \\
		& MLP  &  0.97 (0.01) &   0.96 (0.01) & 0.93 (0.01) & 0.97 (0.01) &  0.91 (0.01) & 0.86 (0.02)  \\
		& CNN  &  \textbf{0.98} (0.01) &  \textbf{0.97} (0.01) & \textbf{0.96} (0.01) & \textbf{0.99} (0.01) &  \textbf{0.91} (0.01) & \textbf{0.88} (0.02)  \\  \hline
		\multirow{3}{*}{\parbox{1.5cm}{Outlier-difference}}
		& MSVR  &2572 (945) &3009 (511) &-1794 (555) &-2038 (688) &13864 (887) &  38828  (2083) \\
		& MLP  &3056 (823) & 3587 (679) & -208  (166) &1417 (403) &14495 (877) & 13526 (2480) \\
		& CNN  &  3060  (809) &3415 (488) & -419 (235) & 1436  (380) & 14503  (883) &12645 (2930) \\ \hline
	\end{tabular}
	\begin{threeparttable}
		\flushleft ~\textbf{Notes.} 
		\begin{tablenotes}
			\small
			\item[a] Each number in the table represents the average value of ten experiments. 
			\item[b] Standard deviations are enclosed in parentheses.
			\item[c] The best PPMCC values achieved by the three machine learning methods are highlighted in boldface. 
			\item[d] A positive outlier-difference means ME produces more outliers than a machine learning method 
                                         while a negative outlier-difference means the machine learning method produces more outliers than ME.
		\end{tablenotes}
	\end{threeparttable}
\end{table}

\begin{table}[!htbp]
	\centering
	\caption{Performance Metric Values of MSVR, MLP and Our CNN Method Based on the Test Set from AR 12665 Collected on 2017 July 13}
	\label{tab:20170713-metrics}
	\begin{tabular}{cl||c||c||c||c||c||c}
		\hline
		&     & $B_{total}$  & \text{$B_x$}  & \text{$B_y$}  & \text{$B_z$} & $\phi$   & $\theta$   \\ \hline
		\multirow{3}{*}{\parbox{1.5cm}{MAE}}
		& MSVR  & 387.23 (9.67) & 582.00 (65.91) & 36.15 (9.34) & 209.48 (21.31) & 23.09 (1.16) & 120.30 (23.96)  \\
		& MLP  & 108.99 (17.69) & 90.04 (5.95) & 76.68 (3.25) & 66.71 (18.89) & 7.67 (0.97) & 23.38 (4.95)  \\
		& CNN  &   87.70 (10.69) &   79.27 (3.60) &  58.04 (3.05) &  53.26 (13.44) &  7.26 (0.80) &  19.94 (4.25)  \\ \hline
		\multirow{3}{*}{\parbox{1.5cm}{PA}} 
		& MSVR  & 19.7\% (1.4\%) & 5.3\% (2.2\%) & 7.8\% (1.8\%) & 78.9\% (1.3\%) & 9.2\% (0.8\%) & 0.5\% (0.5\%) \\
		& MLP  & 87.0\% (2.4\%) & 89.9\% (0.9\%) & 94.5\% (1.1\%) & 91.9\% (1.9\%) & 85.8\% (1.8\%) & 51.2\% (5.7\%) \\
		& CNN  &  90.8\% (1.3\%) &  92.4\% (0.5\%) &  96.4\% (0.8\%) &  93.8\% (1.2\%) &  87.7\% (1.8\%) &  60.0\% (3.7\%) \\  \hline
		\multirow{3}{*}{\parbox{1.5cm}{R-squared}}
		& MSVR  & 0.24 (0.27) & -3.37 (1.53) & -2.39 (0.56) & 0.54 (0.12) & 0.13 (0.09) & -5.67 (3.98)  \\
		& MLP  & 0.85 (0.04) & 0.77 (0.04) & 0.71 (0.06) & 0.86 (0.06) & 0.68 (0.05) & 0.49 (0.12)  \\
		& CNN  &   0.90 (0.02) &   0.80 (0.03) &   0.79 (0.05) &  0.89 (0.04) &  0.70 (0.04) &   0.50 (0.14)  \\  \hline
		\multirow{3}{*}{\parbox{1.5cm}{PPMCC}}
		& MSVR  & 0.73 (0.10) & 0.18 (0.06) & 0.52 (0.08) & 0.84 (0.03) & 0.76 (0.04) & 0.35 (0.14)  \\
		& MLP  & 0.95 (0.01) & 0.89 (0.02) & 0.86 (0.04) & 0.94 (0.02) & 0.84 (0.03) & 0.71 (0.08)  \\
		& CNN  &  \textbf{0.96} (0.01) &  \textbf{0.89} (0.02) & \textbf{0.89} (0.03) & \textbf{0.95} (0.02) &  \textbf{0.85} (0.03) & \textbf{0.72} (0.09)  \\  \hline
		\multirow{3}{*}{\parbox{1.5cm}{Outlier-difference}}
		& MSVR  &5448 (1026) & 6767  (2603) & 3142  (1633) &4052 (864) &34672 (7581) & 93448  (19733) \\
		& MLP  & 5668  (1108) &6623 (2620) &3127 (1674) & 4185  (928) &34716 (7959) &39277 (14562) \\
		& CNN  &5600 (1128) &6267 (2557) &2953 (1583)&4137 (915) & 34721  (7945) &24276 (12194) \\ \hline
	\end{tabular}
	\begin{threeparttable}
		\flushleft ~\textbf{Notes.} 
		\begin{tablenotes}
			\small
			\item[a] Each number in the table represents the average value of ten experiments. 
			\item[b] Standard deviations are enclosed in parentheses.
			\item[c] The best PPMCC values achieved by the three machine learning methods are highlighted in boldface. 
			\item[d] A positive outlier-difference means ME produces more outliers than a machine learning method 
 				while a negative outlier-difference means the machine learning method produces more outliers than ME.
		\end{tablenotes}
	\end{threeparttable}
\end{table}

\begin{table}[!htbp]
	\centering
	\caption{Performance Metric Values of MSVR, MLP and Our CNN Method Based on the Test Set from AR 12673 Collected on 2017 September 6}
	\label{tab:20170906-metrics}
	\begin{tabular}{cl||c||c||c||c||c||c}
		\hline
		&     & $B_{total}$  & \text{$B_x$}  & \text{$B_y$}  & \text{$B_z$} & $\phi$   & $\theta$   \\ \hline
		\multirow{3}{*}{\parbox{1.5cm}{MAE}}
		& MSVR  & 549.84 (0.01) & 851.67 (0.01) & 1079.51 (0.01) & 709.89 (0.01) & 73.19 (0.01) & 56.87 (0.01)  \\
		& MLP  & 339.40 (8.48) & 206.18 (6.98) & 203.56 (5.43) & 223.20 (5.43) & 7.35 (0.14) & 13.23 (0.17)  \\
		& CNN  & 198.92 (3.94) & 150.57 (2.17) & 128.04 (1.63) & 139.30 (4.40) & 5.57 (0.12) & 9.27 (0.20)  \\ \hline
		\multirow{3}{*}{\parbox{1.5cm}{PA}}
		& MSVR  & 17.7\% (0.1\%) & 39.7\% (0.1\%) & 43.9\% (0.1\%) & 6.9\% (0.1\%) & 2.3\% (0.1\%) & 12.9\% (0.1\%)  \\
		& MLP  & 55.9\% (1.3\%) & 70.4\% (1.5\%) & 67.9\% (0.8\%) & 73.5\% (0.8\%) & 82.6\% (1.1\%) & 66.0\% (0.7\%) \\
		& CNN  & 73.6\% (0.4\%) & 80.4\% (0.4\%) & 84.9\% (0.2\%) & 85.7\% (1.5\%) & 90.9\% (0.3\%) & 78.4\% (0.5\%)   \\  \hline
		\multirow{3}{*}{\parbox{1.5cm}{R-squared}}
		& MSVR  & 0.45 (0.01) & -1.37 (0.01) & -7.11 (0.01) & -0.05 (0.01) & -5.49 (0.01) & -1.10 (0.01)  \\
		& MLP  & 0.60 (0.01) & 0.80 (0.01) & 0.57 (0.01) & 0.65 (0.01) & 0.76 (0.01) & 0.77 (0.01)  \\
		& CNN  & 0.84 (0.01) & 0.87 (0.01) & 0.79 (0.01) & 0.74 (0.01) & 0.78 (0.01) & 0.80 (0.01)  \\  \hline
		\multirow{3}{*}{\parbox{1.5cm}{PPMCC}}
		& MSVR  & 0.81 (0.01) & -0.18 (0.01) & -0.31 (0.01) & 0.56 (0.01) & 0.81 (0.01) & 0.54 (0.01)  \\
		& MLP  & 0.85 (0.01) & 0.92 (0.01) & 0.81 (0.01) & 0.82 (0.01) & 0.88 (0.01) & 0.88 (0.01)  \\
		& CNN  & \textbf{0.93} (0.01) & \textbf{0.94} (0.01) & \textbf{0.89} (0.01) & \textbf{0.86} (0.01) & \textbf{0.88} (0.01) & \textbf{0.90} (0.01)  \\  \hline
		\multirow{3}{*}{\parbox{1.5cm}{Outlier-difference}} 
		& MSVR  &19154 (0) &21841 (0)&15980 (0) &12306 (0) &21734 (0) & 32424 (0) \\
		& MLP  &19632 (20) & 22346 (20) & 16780 (38) &12941 (8) &21918 (10) &20692 (552) \\
		& CNN  & 19664 (11) &22234 (46) &16534 (35) & 12965 (7) & 21950 (7) &14294 (1436) \\ \hline
	\end{tabular}
	\begin{threeparttable}
		\flushleft ~\textbf{Notes.} 
		\begin{tablenotes}
			\small
			\item[a] Each number in the table represents the average value of ten experiments. 
			\item[b] Standard deviations are enclosed in parentheses.
			\item[c] The best PPMCC values achieved by the three machine learning methods are highlighted in boldface. 
			\item[d] A positive outlier-difference means ME produces more outliers than a machine learning method 
				while a negative outlier-difference means the machine learning method produces more outliers than ME.
		\end{tablenotes}
	\end{threeparttable}
\end{table}

\subsection{Results of Using Different Active Regions as Training Data}
In the previous subsection we use data points (pixels) from AR 12371 on 2015 June 22 as training data. 
In this subsection we conduct additional experiments by varying training data as follows.
There are four datasets $\text{D}_{1}$, $\text{D}_{2}$, $\text{D}_{3}$, $\text{D}_{4}$ containing the images from
AR 12371 on 2015 June 22, 
AR 12371 on 2015 June 25, 
AR 12665 on 2017 July 13, and
AR 12673 on 2017 September 6 respectively.
In each experiment we randomly select one million pixels (data samples) from one or more datasets to form a training set.
The CNN model is trained on this training set and the trained model is then used to perform Stokes inversion on a test image.
This test image must be from a dataset that is different from those datasets used to construct the training set.
The time point for the test image is
17:33:00 UT on 2015 June 22, 
20:00:00 UT on 2015 June 25, 
18:35:00 UT on 2017 July 13, and
19:18:00 UT on 2017 September 6 respectively.
We use $\text{D}^{\text{train}}_{x}\rightarrow\text{D}^{\text{test}}_{w}$
($\text{D}^{\text{train}}_{x, y}\rightarrow\text{D}^{\text{test}}_{w}$,
$\text{D}^{\text{train}}_{x, y, z}\rightarrow\text{D}^{\text{test}}_{w}$ respectively)
to represent the experiment that uses training data samples
from $\text{D}_{x}$
(training data samples from $\text{D}_{x}$ and $\text{D}_{y}$,
training data samples from $\text{D}_{x}$, $\text{D}_{y}$ and $\text{D}_{z}$ respectively)
and test data samples (pixels) from $\text{D}_{w}$ where $1 \leq x, y, z, w \leq 4$.
Because $\text{D}_{4}$ has only one 720$\times$720 image with 518400 pixels, $\text{D}_{4}$ alone is not used as a training set.
Hence, there are 25 experiments in total.
In each experiment, we calculate the performance metrics MAE, PA,  R-squared, PPMCC and outlier-difference.
Tables \ref{tab:20150622_173300}$-$\ref{tab:20170906_191800} in the Appendix present the experimental results.
Major findings based on these tables are summarized below.
\begin{quote}
	1. Our CNN-inferred results and ME-calculated results are highly correlated and close to each other with a PPMCC of $\sim$0.9 or higher 
          for the total magnetic field strength, regardless of whether the training and test data used by the CNN method are from the same active region (AR) or different ARs, 
          or whether the training and test data are close (e.g., within $\sim$3 days) or distant (e.g., over 2 years) in time. 
          This finding can be seen from Tables \ref{tab:20150622_173300}-\ref{tab:20170906_191800} where the PPMCC of $B_{total}$ in 
        	$\text{D}^{\text{train}}_{2}\rightarrow\text{D}^{\text{test}}_{1}$
	 ($\text{D}^{\text{train}}_{3}\rightarrow\text{D}^{\text{test}}_{1}$,
	$\text{D}^{\text{train}}_{1}\rightarrow\text{D}^{\text{test}}_{2}$,
	$\text{D}^{\text{train}}_{3}\rightarrow\text{D}^{\text{test}}_{2}$,
	$\text{D}^{\text{train}}_{1}\rightarrow\text{D}^{\text{test}}_{3}$,
	$\text{D}^{\text{train}}_{2}\rightarrow\text{D}^{\text{test}}_{3}$,
	$\text{D}^{\text{train}}_{1}\rightarrow\text{D}^{\text{test}}_{4}$,
	$\text{D}^{\text{train}}_{2}\rightarrow\text{D}^{\text{test}}_{4}$, and
	$\text{D}^{\text{train}}_{3}\rightarrow\text{D}^{\text{test}}_{4}$, respectively)
          is 0.956
	(0.924,
	0.983,
	0.951,
	0.976,
	0.979,
	0.936,
	0.927, and
	0.896, respectively).

          2. With respect to the same test image, using the training data from the same AR in which the test image is taken yields a better result with a higher PPMCC
             than using the training and test data that are from different ARs. 
           This finding can be seen from Tables \ref{tab:20150622_173300} and \ref{tab:20150625_200000} where the PPMCC of $B_{total}$ in 
         $\text{D}^{\text{train}}_{2}\rightarrow\text{D}^{\text{test}}_{1}$ is 0.956, which is greater than
         the PPMCC of $B_{total}$, 0.924, in $\text{D}^{\text{train}}_{3}\rightarrow\text{D}^{\text{test}}_{1}$.
        Moreover, the PPMCC of $B_{total}$ in $\text{D}^{\text{train}}_{1}\rightarrow\text{D}^{\text{test}}_{2}$ is 0.983, which is greater than
       the PPMCC of $B_{total}$, 0.951, in $\text{D}^{\text{train}}_{3}\rightarrow\text{D}^{\text{test}}_{2}$.

       3. However, with respect to the same test image, using the training and test data that are close in time does not necessarily yield a better result
           than using the training and test data that are distant in time.
      	This finding can be seen from Table \ref{tab:20170906_191800} where the PPMCC of $B_{total}$ in 
          $\text{D}^{\text{train}}_{1}\rightarrow\text{D}^{\text{test}}_{4}$ is 0.936, which is greater than 
         the PPMCC of $B_{total}$, 0.896, in $\text{D}^{\text{train}}_{3}\rightarrow\text{D}^{\text{test}}_{4}$,
       though $D_3$ is closer to $D_4$ than $D_1$ in time.

	4.  From Tables \ref{tab:20150622_173300}-\ref{tab:20170906_191800}, 
	we can see that the CNN-inferred results have much fewer outliers than the ME-calculated results 
          for all of $B_{total}$, $B_x$, $B_y$, $B_z$, $\phi$, $\theta$  
	in all the experiments 
          except for $B_y$ in Table \ref{tab:20150625_200000}. 
         This finding is consistent with the results reported in Table \ref{tab:single-image-metrics}.
\end{quote}

\section{Discussion and Conclusions}
\label{sec:conclusion}
We develop a new machine learning method to infer vector magnetic fields 
from Stokes profiles of GST/NIRIS based on a convolutional neural network (CNN) and the Milne-Eddington (ME) method.
We then conduct a series of experiments to evaluate the performance of our method.
First, we use data samples (pixels) from AR 12371 collected on 2015 June 22 to train the CNN model 
where the labels (i.e., vector magnetic fields) of the training data samples are calculated by the ME method.
Next, we use the trained model to infer vector magnetic fields from Stokes profiles of pixels in three different unseen test sets.
The first test set contains image data from  AR 12371 collected on 2015 June 25.
The second test set contains image data from AR 12665 collected on 2017 July 13.
The third test set contains image data from AR 12673 collected on 2017 September 6.
We compare our CNN method with the ME method and two related machine learning algorithms, 
multiple support vector regression (MSVR) and multilayer perceptrons (MLP), on the three test sets.
Finally, we conduct more experiments by varying training data to get different trained models and applying the models to different test data.

Our findings based on these experiments are consistent, which are summarized as follows:
\begin{quote}
       1. Our CNN method produces smoother and cleaner magnetic maps with fewer outliers (noise pixels) than the ME method.
       
       2. It takes $\sim$50 seconds for the CNN method to process an image of 720$\times$720 pixels 
comprising Stokes profiles of GST/NIRIS, which is 4$\sim$6 times faster than the current version of the ME method.
       The ability of producing vector magnetic fields in near real-time is essential to space weather forecasting.

     3. Our CNN-inferred results and ME-calculated results are highly correlated and close to each other with a PPMCC of $\sim$0.9 or higher 
          for the total magnetic field strength, regardless of whether the training and test data used by the CNN method are from the same active region (AR) or different ARs, 
          or whether the training and test data are close (e.g., within $\sim$3 days) or distant (e.g., over 2 years) in time. 
	With respect to the same test image, using the training data from the same AR in which the test image is taken yields a better result with a higher PPMCC
             than using the training and test data that are from different ARs. 
       	Hence, for a given test image, it is recommended to adopt the CNN model trained on the same AR from which the test image is collected.
	   
       4.  The CNN-inferred results are closer to the ME-calculated results 
with PPMCC values being closer to 1 on average
than those from the related machine learning methods MSVR and MLP.
In particular, the CNN method outperforms the current best machine learning method (MLP) by 2.6\% on average
in PPMCC. 
This happens because the CNN method is able to exploit the spatial information of the Stokes profiles, 
      and learn latent patterns between the Stokes profiles and ME-calculated vector magnetic fields in a better way. 
\end{quote}

Based on these findings, we conclude that the proposed CNN model can be considered as an alternative, efficient method 
for Stokes inversion for high resolution polarimetric observations obtained by GST/NIRIS.
More accurate and efficient Stokes inversion will improve near real-time prediction of space weather in the future 
as it prepares more accurate magnetic boundary conditions at the solar surface quickly.
With the advent of big and complex observational data gathered from diverse instruments
such as BBSO/GST and the upcoming Daniel K. Inouye Solar Telescope (DKIST),
it is expected that our physics-assisted deep learning-based
CNN tool will be a useful utility for processing and analyzing the data.

We thank the referees for very helpful and thoughtful comments.
The data used in this study were obtained with GST at BBSO, which is operated by New Jersey Institute of Technology. 
Obtaining the excellent data would not have been possible without the help of the BBSO team. 
The BBSO operation is supported by NJIT and NSF grant AGS-1821294. 
The GST operation is partly supported by the Korea Astronomy and Space Science Institute and Seoul National University. 
The related machine learning algorithms studied here were implemented in Python.
This work was supported by NSF grant AGS-1927578.
Y.X., J.J., C.L. and H.W. acknowledge the support of NASA under grants  NNX16AF72G, 80NSSC17K0016, 80NSSC18K0673 and 80NSSC18K1705.

\facilities{Big Bear Solar Observatory}

\clearpage
\appendix
\begin{figure}[H]
	\gridline{\fig{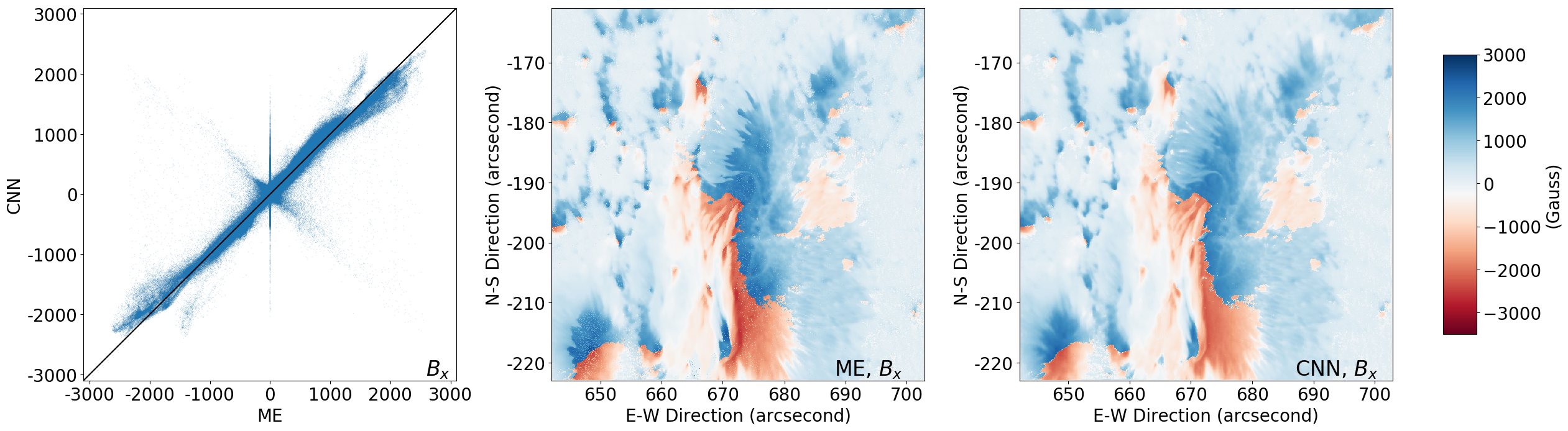}{\textwidth}{}}
	\gridline{\fig{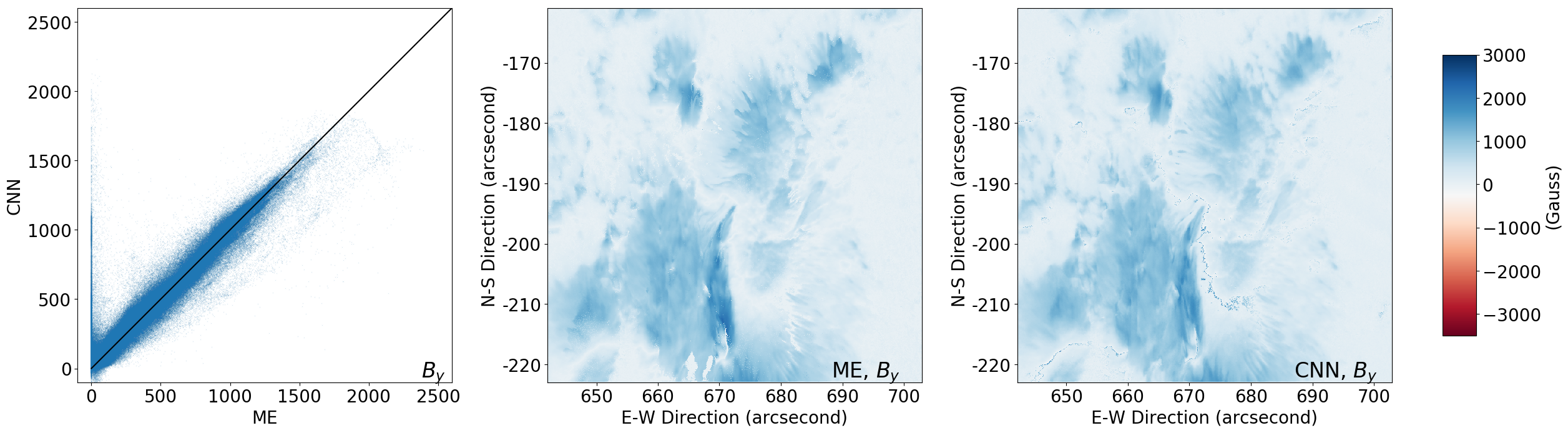}{\textwidth}{}}
	\gridline{\fig{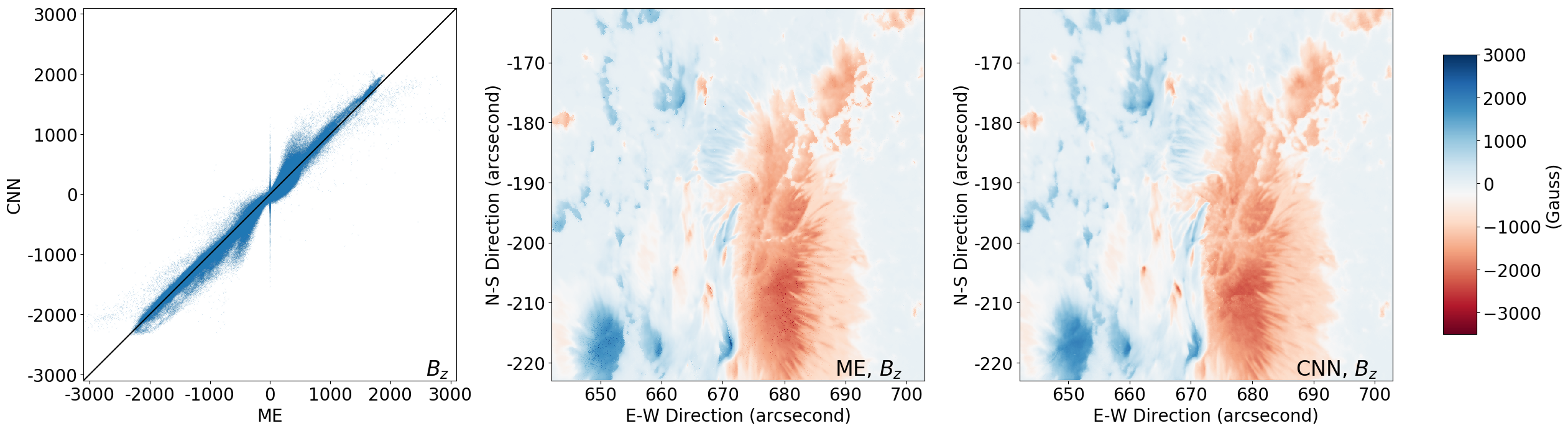}{\textwidth}{}}
	\caption{Comparison between the ME and CNN methods for deriving  $B_{x}$, $B_{y}$ and $B_{z}$ based on the test image from AR 12371 collected on 2015 June 25 20:00:00 UT where training data were taken from the same AR 12371 on 2015 June 22. 
		Displayed from top to bottom are the results for $B_x$, $B_y$ and $B_z$ respectively. 
		The first column shows scatter plots where the X-axis and Y-axis represent the values obtained by the ME and CNN methods respectively. 
		The black diagonal line in each scatter plot corresponds to pixels whose ME-calculated values are identical to CNN-inferred values. 
		The second column shows magnetic maps derived by the ME method. 
		The third column shows magnetic maps inferred by our CNN method. }
	\label{fig:20150625_200000-2} 
\end{figure}

\clearpage
\begin{figure}[H]
	\gridline{\fig{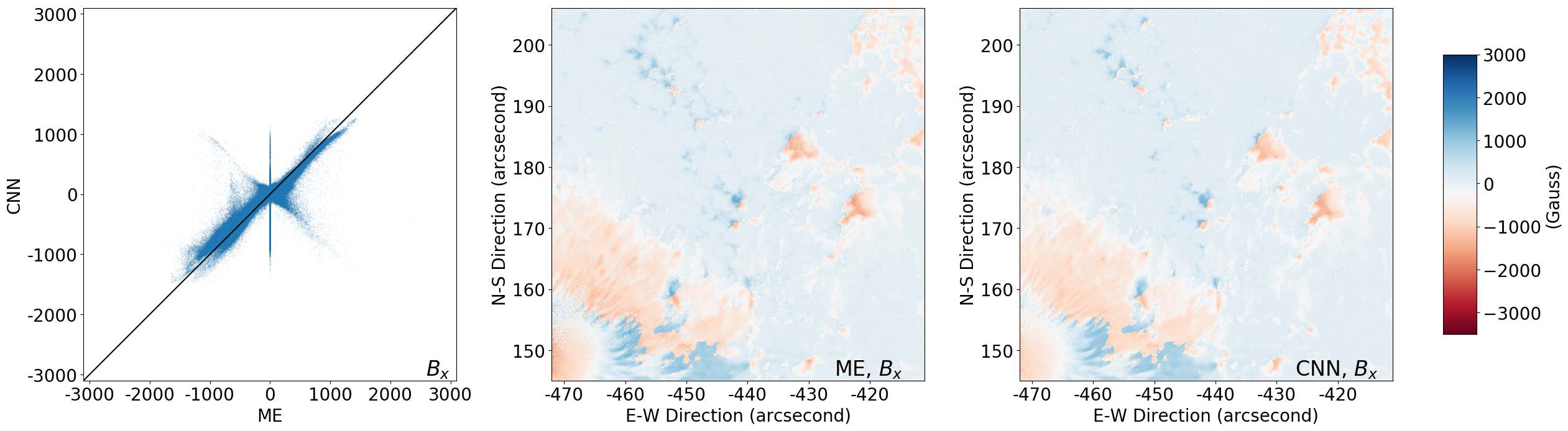}{\textwidth}{}}
	\gridline{\fig{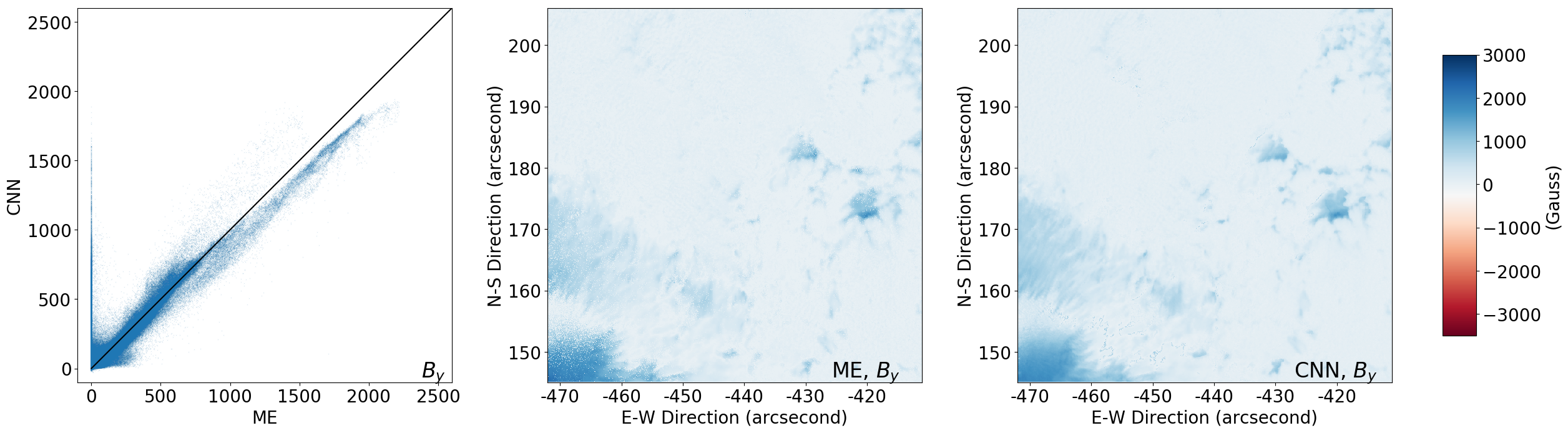}{\textwidth}{}}
	\gridline{\fig{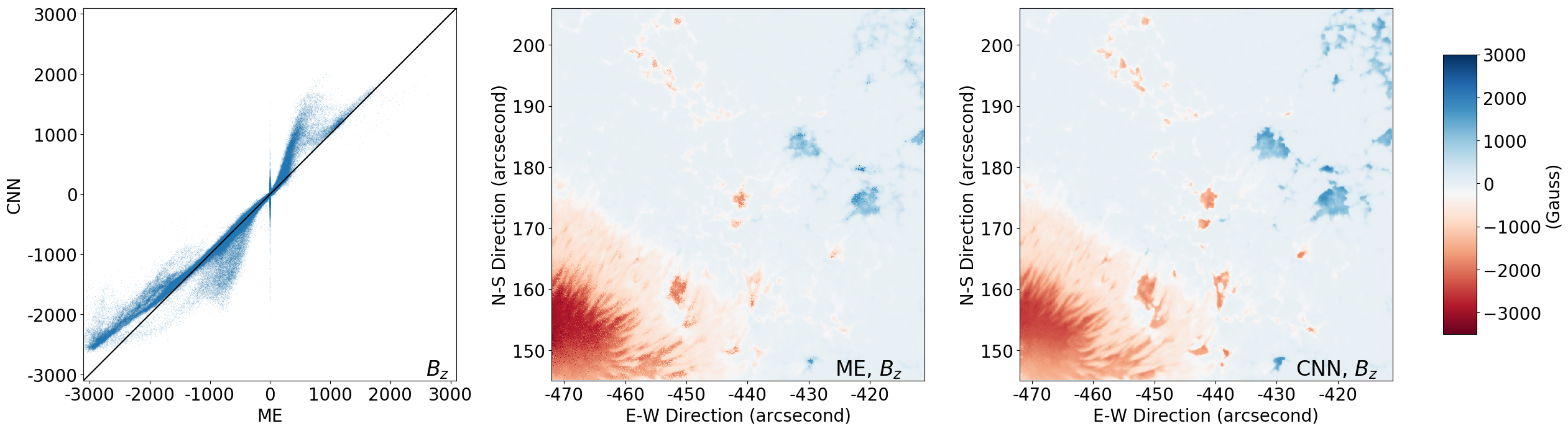}{\textwidth}{}}
	\caption{Comparison between the ME and CNN methods for deriving  $B_{x}$, $B_{y}$ and $B_{z}$ based on the test image from AR 12665 collected on 2017 July 13 18:35:00 UT where training data were taken from AR 12371 on 2015 June 22. 
		Displayed from top to bottom are the results for $B_x$, $B_y$ and $B_z$ respectively. 
		The first column shows scatter plots where the X-axis and Y-axis represent the values obtained by the ME and CNN methods respectively. 
		The black diagonal line in each scatter plot corresponds to pixels whose ME-calculated values are identical to CNN-inferred values. 
		The second column shows magnetic maps derived by the ME method. 
		The third column shows magnetic maps inferred by our CNN method. }
	\label{fig:20170713_183500-2} 
\end{figure}

\clearpage
\begin{figure}[H]
	\gridline{\fig{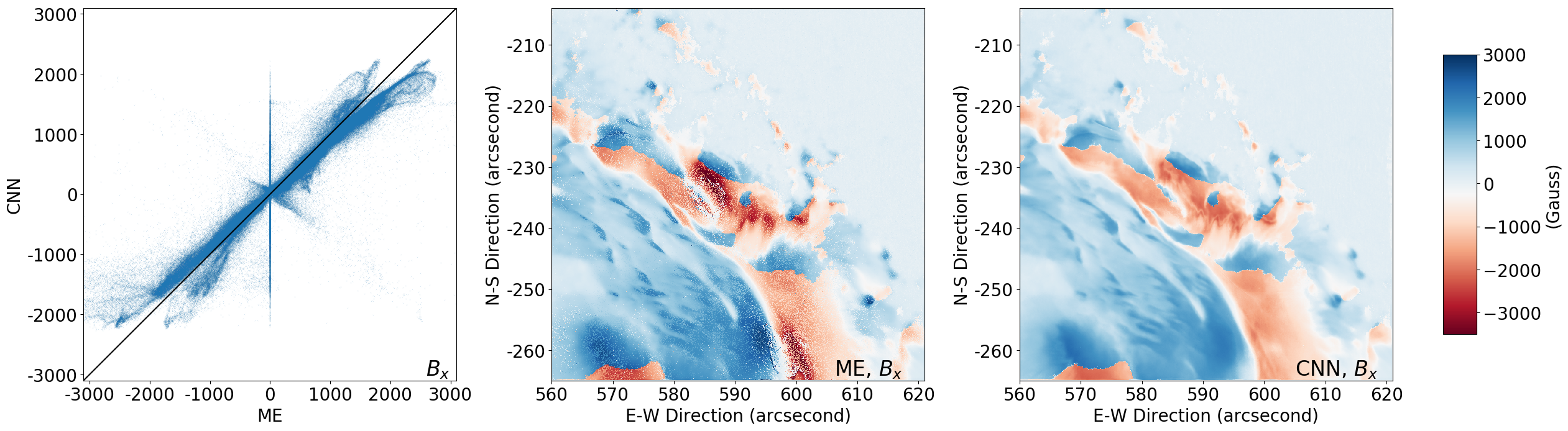}{\textwidth}{}}
	\gridline{\fig{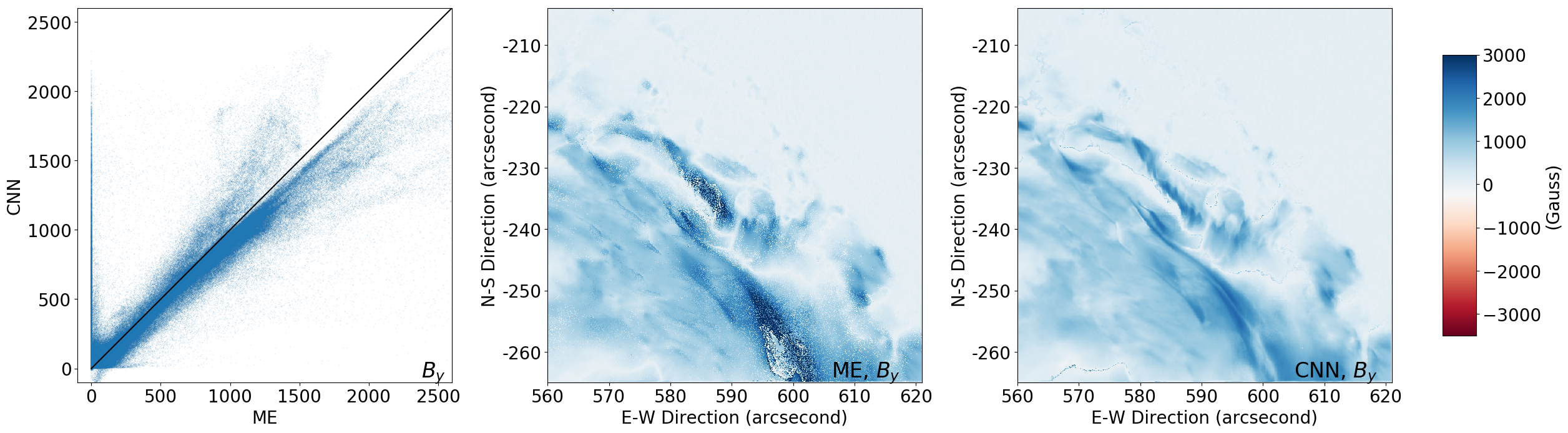}{\textwidth}{}}
	\gridline{\fig{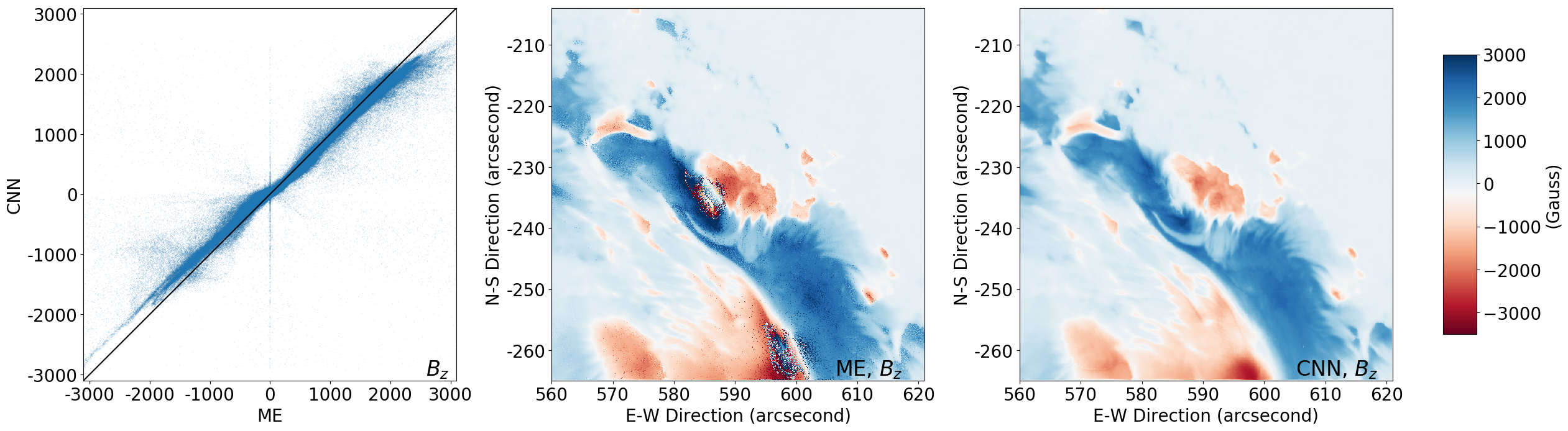}{\textwidth}{}}
	\caption{Comparison between the ME and CNN methods for deriving  $B_{x}$, $B_{y}$ and $B_{z}$ based on the test image from AR 12673 collected on 2017 September 6 19:18:00 UT where training data were taken from AR 12371 on 2015 June 22. 
		Displayed from top to bottom are the results for $B_x$, $B_y$ and $B_z$ respectively. 
		The first column shows scatter plots where the X-axis and Y-axis represent the values obtained by the ME and CNN methods respectively. 
		The black diagonal line in each scatter plot corresponds to pixels whose ME-calculated values are identical to CNN-inferred values. 
		The second column shows magnetic maps derived by the ME method. 
		The third column shows magnetic maps inferred by our CNN method. }
	\label{fig:20170906_191800-2} 
\end{figure}

\clearpage
\begin{table}[!htbp]
	\centering
	\caption{Performance Metric Values of Our CNN Method Based on the Test Images from AR 12371 
            Collected at Ten Different Time Points on 2015 June 25}
	\label{tab:multiple-images-metrics-20150625}
	\begin{tabular}{c||cl||c||c||c||c||c||c}
		\hline
		& & & $B_{total}$  & \text{$B_x$}  & \text{$B_y$}  & \text{$B_z$} & $\phi$   & $\theta$   \\ \hline
		\multirow{40}{*}{2015-06-25 (AR 12371)} 
		& \multirow{5}{*}{17:02:00 UT}
		&MAE & 79.916 & 72.753 & 61.458 & 53.999 & 4.578 & 9.069  \\ 
		&&PA & 91.9\% & 93.4\% &95.3\% &94.6\% &93.6\% &81.3\%  \\ 
		&&R-squared &0.965 &0.935 &0.921 &0.968 &0.823 &0.769  \\ 
		&&PPMCC &0.983 &0.968 &0.961 &0.984 &0.908 &0.879  \\ 
		&&Outlier-difference &3935 & 3881 & 159 & 1857 & 14543 & 18746 \\ \cline{2-9}
		& \multirow{5}{*}{17:20:00 UT}
		&MAE &86.211 &74.157 &56.345 &54.512 &4.254 &8.416  \\ 
		&&PA &90.2\% &92.0\% &96.6\% &94.4\% &94.1\% &81.9\%  \\ 
		&&R-squared &0.961 &0.941 &0.932 &0.966 &0.833 &0.8797  \\ 
		&&PPMCC &0.982 &0.971 &0.968 &0.984 &0.913 &0.895  \\ 
		&&Outlier-difference & 5148 & 3754 & -263 & 2225 &13631 & 14329 \\ \cline{2-9}
		& \multirow{5}{*}{17:41:00 UT}
		&MAE &75.772 &68.146 &57.255 &47.984 &4.184 &8.376  \\ 
		&&PA &92.4\% &93.6\% &96.1\% &95.5\% &93.8\% &82.5\%  \\ 
		&&R-squared &0.968 &0.947 &0.925 &0.974 &0.847 &0.795  \\ 
		&&PPMCC &0.985 &0.974 &0.963 &0.987 &0.920 &0.893  \\ 
		&&Outlier-difference & 2958&2995 &-555 &1596 &12989 &13302 \\ \cline{2-9}
		& \multirow{5}{*}{18:00:00 UT}
		&MAE &77.941 &71.767 &57.802 &50.538 &4.498 &8.586  \\ 
		&&PA &92.6\% &93.6\% &96.7\% &95.4\% &93.9\% &82.4\%  \\ 
		&&R-squared &0.970 &0.943 &0.927 &0.974 &0.825 &0.786  \\ 
		&&PPMCC &0.987 &0.972 &0.966 &0.987 &0.909 &0.888  \\ 
		&&Outlier-difference &2813 &3036 &-630 &1639 &14735 &15265 \\ \cline{2-9}
		& \multirow{5}{*}{18:20:00 UT}
		&MAE &80.294 &74.324 &56.363 &49.600 &4.278 &8.389  \\ 
		&&PA &91.9\% &92.5\% &96.5\% &95.7\% &94.5\% &83.2\%  \\ 
		&&R-squared &0.965 &0.940 &0.931 &0.975 &0.839 &0.788  \\ 
		&&PPMCC &0.984 &0.970 &0.967 &0.988 &0.917 &0.889  \\ 
		&&Outlier-difference &2754 &2840 &-432 &1407 &13432 &12845 \\ \cline{2-9}
		& \multirow{5}{*}{18:40:00 UT}
		&MAE &82.176 &77.420 &57.013 &52.885 &4.760 &8.948  \\ 
		&&PA &91.2\% &92.0\% &96.4\% &94.8\% &93.7\% &82.2\%  \\ 
		&&R-squared &0.964 &0.942 &0.933 &0.972 &0.822 &0.770  \\ 
		&&PPMCC &0.984 &0.971 &0.967 &0.987 &0.907 &0.881  \\ 
		&&Outlier-difference &2508 &2863 &-389 &1253 &15026 &10675 \\ \cline{2-9}
		& \multirow{5}{*}{19:00:00 UT}
		&MAE &79.144 &76.014 &57.960 &52.950 &4.716 &9.507  \\ 
		&&PA &91.7\% &92.8\% &96.1\% &94.9\% &93.7\% &80.8\%  \\ 
		&&R-squared &0.967 &0.943 &0.925 &0.975 &0.828 &0.757  \\ 
		&&PPMCC &0.985 &0.972 &0.964 &0.990 &0.911 &0.872  \\ 
		&&Outlier-difference &2450 &3166 &-424 &1125 &15383 &11646 \\ \cline{2-9}
		& \multirow{5}{*}{19:22:00 UT}
		&MAE &86.917 &79.777 &59.833 &51.695 &4.470 &9.586  \\ 
		&&PA &90.6\% &92.3\% &95.9\% &95.4\% &93.3\% &81.9\%  \\ 
		&&R-squared &0.962 &0.939 &0.924 &0.975 &0.837 &0.742  \\ 
		&&PPMCC &0.983 &0.970 &0.964 &0.988 &0.915 &0.865  \\ &&Outlier-difference &2644 &3508 &-392 &1297 &14243 &11326 \\ \cline{2-9}
		& \multirow{5}{*}{19:41:00 UT}
		&MAE &80.683 &82.218 &58.095 &51.991 &4.775 &11.341  \\ 
		&&PA &91.7\% &91.8\% &96.3\% &95.5\% &93.1\% &78.8\%  \\ 
		&&R-squared &0.966 &0.935 &0.928 &0.974 &0.827 &0.706  \\ 
		&&PPMCC &0.984 &0.968 &0.965 &0.988 &0.910 &0.845  \\ 
		&&Outlier-difference &2426 &3722 &-495 &908 &15939 &11092 \\ \cline{2-9}
		& \multirow{5}{*}{20:00:00 UT} 
		&MAE &  86.660 &  88.997 & 66.140 & 55.653 & 4.867 & 11.136  \\ 
		&&PA &91.6\% &91.3\% &95.2\% &94.7\% &92.2\% &79.1\% \\
		&&R-squared & 0.963 &  0.936 & 0.901 & 0.976 & 0.838 & 0.720 \\   
		&&PPMCC & 0.983 &  0.968 & 0.951 & 0.989 &  0.916 & 0.853 \\  
		&&Outlier-difference &2959 &4380 &-770 &1050 &15108 &7219 \\ \hline
	\end{tabular}
\end{table}

\clearpage
\begin{table}[!htbp]
	\centering
	\caption{Performance Metric Values of Our CNN Method Based on the Test Images from AR 12665 
         Collected at Ten Different Time Points on 2017 July 13}
	\label{tab:multiple-images-metrics-20170713}
	\begin{tabular}{c||cl||c||c||c||c||c||c}
		\hline
		& & & $B_{total}$  & \text{$B_x$}  & \text{$B_y$}  & \text{$B_z$} & $\phi$   & $\theta$   \\ \hline
		\multirow{40}{*}{2017-07-13 (AR 12665)} 
		& \multirow{5}{*}{17:18:00 UT}
		&MAE &96.763 &77.228 &59.555 &65.277 &6.995 &15.623  \\ 
		&&PA &89.9\% &93.5\% &95.7\% &92.7\% &88.5\% &62.2\%  \\ 
		&&R-squared &0.895 &0.796 &0.726 &0.875 &0.716 &0.686  \\ 
		&&PPMCC &0.961 &0.893 &0.857 &0.947 &0.851 &0.834  \\ 
		&&Outlier-difference &5612 &8931 &5341 &4805 &36440 &39816 \\ \cline{2-9}
		& \multirow{5}{*}{17:54:00 UT}
		&MAE &108.101 &86.635 &60.292 &83.230 &8.276 &16.899  \\ 
		&&PA &90.8\% &92.3\% &95.6\% &92.8\% &86.0\% &60.8\%  \\ 
		&&R-squared &0.866 &0.745 &0.695 &0.789 &0.647 &0.677  \\ 
		&&PPMCC &0.953 &0.864 &0.838 &0.902 &0.814 &0.829  \\ 
		&&Outlier-difference &5430 &11516 &5702 &5728 &45541 &41897 \\ \cline{2-9}
		& \multirow{5}{*}{18:25:00 UT}
		&MAE &95.509 &81.222 &59.119 &66.639 &7.984 &17.809  \\ 
		&&PA &89.6\% &92.2\% &95.7\% &91.9\% &86.5\% &60.1\%  \\ 
		&&R-squared &0.914 &0.822 &0.792 &0.874 &0.661 &0.664  \\ 
		&&PPMCC &0.971 &0.907 &0.893 &0.947 &0.824 &0.820  \\ 
		&&Outlier-difference &3874 &8158 &4134 &3657 &42169 &39937 \\ \cline{2-9}
		&\multirow{5}{*}{18:35:00 UT}
		&MAE & 73.684 &  71.555 & 51.170 & 49.023 & 7.573 & 17.437  \\ 
		&&PA &91.5\% &93.3\% &96.4\% &92.6\% &84.8\% &60.6\% \\
		&&R-squared & 0.950 &  0.841 & 0.851 & 0.941 & 0.663 & 0.665 \\ 
		&&PPMCC & 0.976 &  0.918 & 0.926 & 0.971 &  0.827 & 0.821 \\  
		&&Outlier-difference &3801 &7280 &3413 &2478 &35649 &28274 \\ \cline{2-9}
		& \multirow{5}{*}{20:19:00 UT}
		&MAE &75.811 &78.550 &55.695 &38.701 &8.014 &26.263  \\ 
		&&PA &92.8\% &92.1\% &97.4\% &95.7\% &86.6\% &55.6\%  \\ 
		&&R-squared &0.915 &0.826&0.831 &0.930 &0.680 &0.456  \\ 
		&&PPMCC &0.960 &0.910 &0.916 &0.966 &0.831 &0.693  \\ 
		&&Outlier-difference &5089 &4479 &1900 &3341 &41315 & 12610 \\ \cline{2-9}
		& \multirow{5}{*}{20:52:00 UT}
		&MAE &77.201 &78.757 &56.624 &41.618 &7.979 &26.759  \\ 
		&&PA &92.6\% &92.1\% &97.6\% &95.1\% &86.4\% &54.3\%  \\ 
		&&R-squared &0.914 &0.805 &0.827 &0.926 &0.682 &0.401  \\ 
		&&PPMCC &0.957 &0.897 &0.913 &0.963 &0.834 &0.656  \\ 
		&&Outlier-difference &5878 &4499 &1788 &4019 &38418 & 10550 \\ \cline{2-9}
		& \multirow{5}{*}{21:20:00 UT}
		&MAE &80.011 &77.987 &57.099 &42.847 &7.200 &25.012  \\ 
		&&PA &92.2\% &92.4\% &97.4\% & 94.9\% &88.3\% &55.1\%  \\ 
		&&R-squared &0.901 &0.799 &0.805 &0.918 &0.710 &0.352  \\ 
		&&PPMCC &0.952 &0.895 &0.901 &0.961 &0.851 &0.624  \\ 
		&&Outlier-difference &5997 &4505 &1771 &3992 &33783 &15985 \\ \cline{2-9}
		& \multirow{5}{*}{21:48:00 UT}
		&MAE &84.746 &78.946 &57.296 &43.822 &6.471 &21.160  \\ 
		&&PA &89.6\% &91.9\% &97.0\% &94.9\% &90.1\% &60.7\%  \\ 
		&&R-squared &0.895 &0.791 &0.808 &0.917 &0.728 &0.380  \\
		&&PPMCC &0.953 &0.891 &0.901 &0.963 &0.859 &0.643  \\  
		&&Outlier-difference &6234 &4556 &1938 &3916 &28967 &19769 \\ \cline{2-9}
		& \multirow{5}{*}{22:18:00 UT}
		&MAE &95.769 &81.151 &61.962 &52.672 &5.890 &16.784  \\ 
		&&PA &88.8\% &91.7\% &95.7\% &93.6\% &90.3\% &64.6\%  \\ 
		&&R-squared &0.869 &0.771 &0.779 &0.893 &0.776 &0.364  \\ 
		&&PPMCC &0.942 &0.881 &0.884 &0.952 &0.888 &0.640  \\ 
		&&Outlier-difference &7325 &4238 &1721 &5022 &22740 & 15454\\ \cline{2-9}
		& \multirow{5}{*}{22:39:00 UT}
		&MAE &89.352 &80.647 &61.617 &48.760 &6.226 &15.683  \\ 
		&&PA &90.3\% &92.0\% &95.8\% &94.4\% &89.4\% &65.9\%  \\ 
		&&R-squared &0.889 &0.774 &0.751 &0.913 &0.775 &0.399  \\ 
		&&PPMCC &0.951 &0.885 &0.868 &0.961 &0.889 &0.664  \\ 
		&&Outlier-difference &6757 &4506 &1826 &4408 &22186 &18471 \\ \hline
	\end{tabular}
\end{table}

\clearpage
\begin{table}[!htbp]
	\centering
	\caption{Performance Metric Values of Our CNN Method Obtained by Using 
				$\text{D}_{1}$ to Form Test Data and
				Different Combinations of $\text{D}_{2}$, $\text{D}_{3}$, $\text{D}_{4}$ to Form Training Data}
	\label{tab:20150622_173300}
	\begin{tabular}{cl||c||c||c||c||c||c}
		\hline
		&  & $B_{total}$  & \text{$B_x$}  & \text{$B_y$}  & \text{$B_z$} & $\phi$   & $\theta$   \\ \hline
		\multirow{6}{*}{MAE}
		&$\text{D}^{\text{train}}_{2}\rightarrow\text{D}^{\text{test}}_{1}$ &112.104 &70.871 &77.554 &83.761 &5.040 & 10.286 \\
		&$\text{D}^{\text{train}}_{3}\rightarrow\text{D}^{\text{test}}_{1}$ &168.905 &96.727 &116.505 &112.322 &5.724 &11.874 \\
		&$\text{D}^{\text{train}}_{2, 3}\rightarrow\text{D}^{\text{test}}_{1}$ &99.187 &74.859 &75.777 &81.588 &5.330 &10.330   \\
		&$\text{D}^{\text{train}}_{2, 4}\rightarrow\text{D}^{\text{test}}_{1}$ &96.981 &78.739 &77.672 &70.458 &5.111 &11.356   \\
		&$\text{D}^{\text{train}}_{3, 4}\rightarrow\text{D}^{\text{test}}_{1}$ &137.511 &87.619 &105.794 &83.560 &5.315 &11.086 \\
		&$\text{D}^{\text{train}}_{2, 3, 4}\rightarrow\text{D}^{\text{test}}_{1}$ &97.594 &73.092 &75.116 &74.675 &5.095 &10.258 \\ \hline
		\multirow{6}{*}{PA} 
		&$\text{D}^{\text{train}}_{2}\rightarrow\text{D}^{\text{test}}_{1}$ &88.5\% &93.6\% &92.4\% &90.8\% &90.6\% &78.0\% \\
		&$\text{D}^{\text{train}}_{3}\rightarrow\text{D}^{\text{test}}_{1}$ &71.8\% &89.5\% &80.8\% &85.7\% &89.0\% &78.6\% \\
		&$\text{D}^{\text{train}}_{2, 3}\rightarrow\text{D}^{\text{test}}_{1}$ &89.2\% &92.7\% &92.5\% &91.2\% &89.6\% &78.5\%   \\
		&$\text{D}^{\text{train}}_{2, 4}\rightarrow\text{D}^{\text{test}}_{1}$ &90.0\% &92.2\% &92.6\% &92.4\% &90.5\% &76.3\%  \\
		&$\text{D}^{\text{train}}_{3, 4}\rightarrow\text{D}^{\text{test}}_{1}$ &81.7\% &91.3\% &87.6\% &90.5\% &90.2\% &78.9\%   \\
		&$\text{D}^{\text{train}}_{2, 3, 4}\rightarrow\text{D}^{\text{test}}_{1}$ &89.7\% &92.7\% &92.5\% &92.3\% &90.2\% &79.1\%   \\ \hline
		\multirow{6}{*}{R-squared} 
		&$\text{D}^{\text{train}}_{2}\rightarrow\text{D}^{\text{test}}_{1}$ &0.903 &0.913 &0.878 &0.955 &0.867 &0.710  \\
		&$\text{D}^{\text{train}}_{3}\rightarrow\text{D}^{\text{test}}_{1}$ &0.845 &0.860 &0.810 &0.929 &0.867 &0.576   \\
		&$\text{D}^{\text{train}}_{2, 3}\rightarrow\text{D}^{\text{test}}_{1}$ &0.907 &0.910 &0.875 &0.953 &0.868 &0.706   \\
		&$\text{D}^{\text{train}}_{2, 4}\rightarrow\text{D}^{\text{test}}_{1}$ &0.909 &0.899 &0.862 &0.962 &0.861 &0.657   \\
		&$\text{D}^{\text{train}}_{3, 4}\rightarrow\text{D}^{\text{test}}_{1}$ &0.886 &0.888 &0.830 &0.954 &0.864 &0.661   \\
		&$\text{D}^{\text{train}}_{2, 3, 4}\rightarrow\text{D}^{\text{test}}_{1}$ &0.904 &0.908 &0.874 &0.956 &0.869 &0.701 \\ \hline
		\multirow{6}{*}{PPMCC} 
		&$\text{D}^{\text{train}}_{2}\rightarrow\text{D}^{\text{test}}_{1}$ &0.956 &0.956 &0.937 &0.982 &0.935 &0.847  \\
		&$\text{D}^{\text{train}}_{3}\rightarrow\text{D}^{\text{test}}_{1}$&0.924 &0.933 &0.929 &0.965 &0.933 &0.780 \\
		&$\text{D}^{\text{train}}_{2, 3}\rightarrow\text{D}^{\text{test}}_{1}$ &0.954 &0.956 &0.936 &0.980 &0.936 &0.846  \\
		&$\text{D}^{\text{train}}_{2, 4}\rightarrow\text{D}^{\text{test}}_{1}$ &0.954 &0.951 &0.932 &0.982 &0.932 &0.822   \\
		&$\text{D}^{\text{train}}_{3, 4}\rightarrow\text{D}^{\text{test}}_{1}$ &0.946 &0.947 &0.931 &0.978 &0.932 &0.821   \\
		&$\text{D}^{\text{train}}_{2, 3, 4}\rightarrow\text{D}^{\text{test}}_{1}$ &0.952 &0.955 &0.935 &0.980 &0.936 &0.843 \\ \hline
		\multirow{6}{*}{Outlier-difference} 
		& $\text{D}^{\text{train}}_{2}\rightarrow\text{D}^{\text{test}}_{1}$ &9396 &4718 &3419 &6808 &33687 &22528  \\
		& $\text{D}^{\text{train}}_{3}\rightarrow\text{D}^{\text{test}}_{1}$ &9527 &3625 &3355 &6554 &33202 &27896   \\
		& $\text{D}^{\text{train}}_{2, 3}\rightarrow\text{D}^{\text{test}}_{1}$ &9266 &4045 &3120 &6760 &33114 &25720   \\
		& $\text{D}^{\text{train}}_{2, 4}\rightarrow\text{D}^{\text{test}}_{1}$ &9185 &4480 &3086 &6782 &33754 &24704   \\
		& $\text{D}^{\text{train}}_{3, 4}\rightarrow\text{D}^{\text{test}}_{1}$ &9160 &3921 &2878 &6771 &33671 &33627   \\
		& $\text{D}^{\text{train}}_{2, 3, 4}\rightarrow\text{D}^{\text{test}}_{1}$ &8668 &4528 &3301 &6813 &33691 &26670 \\ \hline
	\end{tabular}
\end{table}

\clearpage
\begin{table}[!htbp]
	\centering
	\caption{Performance Metric Values of Our CNN Method Obtained by Using 
				$\text{D}_{2}$ to Form Test Data and
				Different Combinations of $\text{D}_{1}$, $\text{D}_{3}$, $\text{D}_{4}$ to Form Training Data}
	\label{tab:20150625_200000}
	\begin{tabular}{cl||c||c||c||c||c||c}
		\hline
		&  & $B_{total}$  & \text{$B_x$}  & \text{$B_y$}  & \text{$B_z$} & $\phi$   & $\theta$   \\ \hline
		\multirow{6}{*}{MAE}
		&$\text{D}^{\text{train}}_{1}\rightarrow\text{D}^{\text{test}}_{2}$ &86.660 &88.997 &66.140 &55.653 &4.867 &11.136  \\
		&$\text{D}^{\text{train}}_{3}\rightarrow\text{D}^{\text{test}}_{2}$ &165.558 &132.399 &92.024 &109.629 &6.157 &12.191 \\
		&$\text{D}^{\text{train}}_{1, 3}\rightarrow\text{D}^{\text{test}}_{2}$ &83.631 &86.654 &61.893 &51.414 &4.650 &10.949  \\
		&$\text{D}^{\text{train}}_{1, 4}\rightarrow\text{D}^{\text{test}}_{2}$ &90.098 &88.494 &63.458 &60.282 &4.935 &10.595   \\
		&$\text{D}^{\text{train}}_{3, 4}\rightarrow\text{D}^{\text{test}}_{2}$ &133.132 &104.756 &85.091 &79.925 &5.250 &11.805 \\
		&$\text{D}^{\text{train}}_{1, 3, 4}\rightarrow\text{D}^{\text{test}}_{2}$ &79.830 &84.662 &59.412 &50.448 &4.736 &11.023 \\ \hline
		\multirow{6}{*}{PA} 
		&$\text{D}^{\text{train}}_{1}\rightarrow\text{D}^{\text{test}}_{2}$ &91.6\% &91.3\% &95.2\% &94.7\% &92.2\% &79.1\% \\
		&$\text{D}^{\text{train}}_{3}\rightarrow\text{D}^{\text{test}}_{2}$ &69.4\% &81.8\% &87.8\% &81.2\% &87.3\% &74.9\% \\
		&$\text{D}^{\text{train}}_{1, 3}\rightarrow\text{D}^{\text{test}}_{2}$ &89.7\% &90.5\% &95.5\% &95.3\% &93.3\% &79.2\%  \\
		&$\text{D}^{\text{train}}_{1, 4}\rightarrow\text{D}^{\text{test}}_{2}$ &89.0\% &90.5\% &95.1\% &94.0\% &92.6\% &80.2\%   \\
		&$\text{D}^{\text{train}}_{3, 4}\rightarrow\text{D}^{\text{test}}_{2}$ &79.0\% &88.1\% &89.7\% &89.7\% &92.3\% &76.8\%   \\
		&$\text{D}^{\text{train}}_{1, 3, 4}\rightarrow\text{D}^{\text{test}}_{2}$ &91.9\% &92.0\% &95.7\% &96.0\% &93.1\% &78.8\%  \\ \hline
		\multirow{6}{*}{R-squared} 
		& $\text{D}^{\text{train}}_{1}\rightarrow\text{D}^{\text{test}}_{2}$ &0.963 &0.936 &0.901 &0.976 &0.838 &0.720  \\
		& $\text{D}^{\text{train}}_{3}\rightarrow\text{D}^{\text{test}}_{2}$ &0.893 &0.899 &0.850 &0.928 &0.828 &0.695   \\
		& $\text{D}^{\text{train}}_{1, 3}\rightarrow\text{D}^{\text{test}}_{2}$ &0.962 &0.937 &0.914 &0.979 &0.844 &0.724   \\
		& $\text{D}^{\text{train}}_{1, 4}\rightarrow\text{D}^{\text{test}}_{2}$ &0.956 &0.936 &0.907 &0.972 &0.839 &0.727   \\
		& $\text{D}^{\text{train}}_{3, 4}\rightarrow\text{D}^{\text{test}}_{2}$ &0.937 &0.927 &0.858 &0.964 &0.837 &0.711   \\
		& $\text{D}^{\text{train}}_{1, 3, 4}\rightarrow\text{D}^{\text{test}}_{2}$ &0.966 &0.938 &0.918 &0.980 &0.842 &0.724 \\ \hline
		\multirow{6}{*}{PPMCC} 
		&$\text{D}^{\text{train}}_{1}\rightarrow\text{D}^{\text{test}}_{2}$ &0.983 &0.968 &0.951 &0.989 &0.916 &0.853  \\
		&$\text{D}^{\text{train}}_{3}\rightarrow\text{D}^{\text{test}}_{2}$ &0.951 &0.949 &0.939 &0.982 &0.915 &0.846 \\
		&$\text{D}^{\text{train}}_{1, 3}\rightarrow\text{D}^{\text{test}}_{2}$ &0.982 &0.968 &0.957 &0.990 &0.919 &0.855  \\
		&$\text{D}^{\text{train}}_{1, 4}\rightarrow\text{D}^{\text{test}}_{2}$ &0.981 &0.968 &0.955 &0.988 &0.916 &0.856   \\
		&$\text{D}^{\text{train}}_{3, 4}\rightarrow\text{D}^{\text{test}}_{2}$ &0.982 &0.968 &0.954 &0.989 &0.916 &0.850   \\
		&$\text{D}^{\text{train}}_{1, 3, 4}\rightarrow\text{D}^{\text{test}}_{2}$ &0.984 &0.969 &0.960 &0.990 &0.918 &0.855   \\ \hline
		\multirow{6}{*}{Outlier-difference} 
		&$\text{D}^{\text{train}}_{1}\rightarrow\text{D}^{\text{test}}_{2}$ &2959 &4380 &-770 &1050 &15108 &7219  \\
		&$\text{D}^{\text{train}}_{3}\rightarrow\text{D}^{\text{test}}_{2}$ &2950 &3904 &-246 &1032 &15054 &10038   \\
		&$\text{D}^{\text{train}}_{1, 3}\rightarrow\text{D}^{\text{test}}_{2}$ &2948 &4354 &-666 &1053 &15108 &7392   \\
		&$\text{D}^{\text{train}}_{1, 4}\rightarrow\text{D}^{\text{test}}_{2}$ &2954 &4231 &-574 &1055 &15108 &11235   \\
		&$\text{D}^{\text{train}}_{3, 4}\rightarrow\text{D}^{\text{test}}_{2}$ &2953 &3926 &-533 &1057 &15000 &9161   \\
		&$\text{D}^{\text{train}}_{1, 3, 4}\rightarrow\text{D}^{\text{test}}_{2}$ &2959 &4380 &-631 &1053 &15108 &9410   \\ \hline
	\end{tabular}
\end{table}

\clearpage
\begin{table}[!htbp]
	\centering
	\caption{Performance Metric Values of Our CNN Method Obtained by Using 
				$\text{D}_{3}$ to Form Test Data and
				Different Combinations of $\text{D}_{1}$, $\text{D}_{2}$, $\text{D}_{4}$ to Form Training Data}
	\label{tab:20170713_183500}
	\begin{tabular}{cl||c||c||c||c||c||c}
		\hline
		&  & $B_{total}$  & \text{$B_x$}  & \text{$B_y$}  & \text{$B_z$} & $\phi$   & $\theta$   \\ \hline
		\multirow{6}{*}{MAE}
		&$\text{D}^{\text{train}}_{1}\rightarrow\text{D}^{\text{test}}_{3}$ &73.683 &71.555 &51.170 &49.023 &7.573 &17.437  \\
		&$\text{D}^{\text{train}}_{2}\rightarrow\text{D}^{\text{test}}_{3}$ &98.412 &83.441 &55.674 &58.326 &7.330 &18.232  \\
		&$\text{D}^{\text{train}}_{1, 2}\rightarrow\text{D}^{\text{test}}_{3}$ &70.574 &68.776 &48.467 &43.919 &7.381 &16.780  \\
		&$\text{D}^{\text{train}}_{1, 4}\rightarrow\text{D}^{\text{test}}_{3}$ &68.492 &66.340 &48.593 &48.398 &7.394 &15.661  \\
		&$\text{D}^{\text{train}}_{2, 4}\rightarrow\text{D}^{\text{test}}_{3}$ &68.903 &64.068 &44.475 &46.860 &6.539 &14.911  \\
		&$\text{D}^{\text{train}}_{1, 2, 4}\rightarrow\text{D}^{\text{test}}_{3}$ &68.876 &67.419 &48.227 &43.239 &7.250 &16.839  \\ \hline
		\multirow{6}{*}{PA} 
		&$\text{D}^{\text{train}}_{1}\rightarrow\text{D}^{\text{test}}_{3}$ &91.5\% &93.3\% &96.4\% &92.6\% &84.8\% &60.6\% \\
		&$\text{D}^{\text{train}}_{2}\rightarrow\text{D}^{\text{test}}_{3}$ &90.1\% &92.3\% &95.7\% &93.9\% &87.0\% &54.8\% \\
		&$\text{D}^{\text{train}}_{1, 2}\rightarrow\text{D}^{\text{test}}_{3}$ &92.8\% &93.8\% &96.6\% &94.1\% &85.6\% &61.3\% \\
		&$\text{D}^{\text{train}}_{1, 4}\rightarrow\text{D}^{\text{test}}_{3}$ &91.1\% &93.1\% &95.8\% &93.5\% &84.2\% &65.7\%  \\
		&$\text{D}^{\text{train}}_{2, 4}\rightarrow\text{D}^{\text{test}}_{3}$ &93.1\% &93.7\% &96.7\% &93.8\% &87.1\% &68.0\%  \\
		&$\text{D}^{\text{train}}_{1, 2, 4}\rightarrow\text{D}^{\text{test}}_{3}$ &91.5\% &93.9\% &96.7\% &93.1\% &86.2\% &61.9\%  \\ \hline
		\multirow{6}{*}{R-squared} 
		&$\text{D}^{\text{train}}_{1}\rightarrow\text{D}^{\text{test}}_{3}$ &0.950 &0.841 &0.851 &0.941 &0.663 &0.665  \\
		&$\text{D}^{\text{train}}_{2}\rightarrow\text{D}^{\text{test}}_{3}$ &0.926 &0.830 &0.850 &0.924 &0.661 &0.678  \\
		&$\text{D}^{\text{train}}_{1, 2}\rightarrow\text{D}^{\text{test}}_{3}$ &0.957 &0.849 &0.857 &0.955 &0.665 &0.688  \\
		&$\text{D}^{\text{train}}_{1, 4}\rightarrow\text{D}^{\text{test}}_{3}$ &0.951 &0.848 &0.855 &0.944 &0.667 &0.698  \\
		&$\text{D}^{\text{train}}_{2, 4}\rightarrow\text{D}^{\text{test}}_{3}$ &0.952 &0.845 &0.860 &0.946 &0.703 &0.696  \\
		&$\text{D}^{\text{train}}_{1, 2, 4}\rightarrow\text{D}^{\text{test}}_{3}$ &0.959 &0.848 &0.858 &0.956 &0.672 &0.680  \\ \hline
		\multirow{6}{*}{PPMCC} 
		&$\text{D}^{\text{train}}_{1}\rightarrow\text{D}^{\text{test}}_{3}$ &0.976 &0.918 &0.926 &0.971 &0.827 &0.821  \\
		&$\text{D}^{\text{train}}_{2}\rightarrow\text{D}^{\text{test}}_{3}$ &0.979 &0.913 &0.923 &0.978 &0.829 &0.829  \\
		&$\text{D}^{\text{train}}_{1, 2}\rightarrow\text{D}^{\text{test}}_{3}$ &0.979 &0.921 &0.928 &0.978 &0.828 &0.834  \\
		&$\text{D}^{\text{train}}_{1, 4}\rightarrow\text{D}^{\text{test}}_{3}$ &0.975 &0.921 &0.926 &0.973 &0.828 &0.839  \\
		&$\text{D}^{\text{train}}_{2, 4}\rightarrow\text{D}^{\text{test}}_{3}$ &0.976 &0.922 &0.929 &0.973 &0.843 &0.844  \\
		&$\text{D}^{\text{train}}_{1, 2, 4}\rightarrow\text{D}^{\text{test}}_{3}$ &0.980 &0.921 &0.928 &0.978 &0.831 &0.830  \\ \hline
		\multirow{6}{*}{Outlier-difference} 
		&$\text{D}^{\text{train}}_{1}\rightarrow\text{D}^{\text{test}}_{3}$ &3801 &7280 &3413 &2478 &35649 &28274  \\
		&$\text{D}^{\text{train}}_{2}\rightarrow\text{D}^{\text{test}}_{3}$ &3837 &7284 &3545 &2497 &35661 &27447  \\
		&$\text{D}^{\text{train}}_{1, 2}\rightarrow\text{D}^{\text{test}}_{3}$ &3800 &7252 &3433 &2480 &35647 &25888  \\
		&$\text{D}^{\text{train}}_{1, 4}\rightarrow\text{D}^{\text{test}}_{3}$ &3812 &7200 &3405 &2481 &35645 &31320  \\
		&$\text{D}^{\text{train}}_{2, 4}\rightarrow\text{D}^{\text{test}}_{3}$ &3824 &6873 &3494 &2496 &35657 &26492  \\
		&$\text{D}^{\text{train}}_{1, 2, 4}\rightarrow\text{D}^{\text{test}}_{3}$ &3801 &7371 &3463 &2490 &35645 &24124  \\ \hline
	\end{tabular}
\end{table}

\clearpage
\begin{table}[!htbp]
	\centering
	\caption{Performance Metric Values of Our CNN Method Obtained by Using 
				$\text{D}_{4}$ to Form Test Data and
				Different Combinations of $\text{D}_{1}$, $\text{D}_{2}$, $\text{D}_{3}$ to Form Training Data}
	\label{tab:20170906_191800}
	\begin{tabular}{cl||c||c||c||c||c||c}
		\hline
		&  & $B_{total}$  & \text{$B_x$}  & \text{$B_y$}  & \text{$B_z$} & $\phi$   & $\theta$   \\ \hline
		\multirow{7}{*}{MAE}
		&$\text{D}^{\text{train}}_{1}\rightarrow\text{D}^{\text{test}}_{4}$ &193.680 &146.010 &124.783 &136.892 &5.497 &9.009  \\
		&$\text{D}^{\text{train}}_{2}\rightarrow\text{D}^{\text{test}}_{4}$ &246.086 &160.538 &131.986 &186.657 &6.296 &9.501  \\
		&$\text{D}^{\text{train}}_{3}\rightarrow\text{D}^{\text{test}}_{4}$ &231.481 &153.664 &129.813 &173.582 &5.823 &7.473  \\
		&$\text{D}^{\text{train}}_{1, 2}\rightarrow\text{D}^{\text{test}}_{4}$ &198.832 &143.087 &123.287 &146.410 &5.363 &8.729 \\
		&$\text{D}^{\text{train}}_{1, 3}\rightarrow\text{D}^{\text{test}}_{4}$ &204.086 &143.244 &123.685 &148.227 &5.284 &7.925  \\
		&$\text{D}^{\text{train}}_{2, 3}\rightarrow\text{D}^{\text{test}}_{4}$ &201.117 &137.369 &119.157 &162.063 &5.713 &7.577  \\
		&$\text{D}^{\text{train}}_{1, 2, 3}\rightarrow\text{D}^{\text{test}}_{4}$ &207.075 &148.718 &127.467 &146.775 &5.674 &8.679  \\ \hline
		\multirow{7}{*}{PA} 
		&$\text{D}^{\text{train}}_{1}\rightarrow\text{D}^{\text{test}}_{4}$ &75.0\% &80.1\% &86.2\% &87.2\% &91.3\% &79.1\% \\
		&$\text{D}^{\text{train}}_{2}\rightarrow\text{D}^{\text{test}}_{4}$ &54.9\% &77.6\% &83.7\% &77.0\% &87.7\% &76.3\% \\
		&$\text{D}^{\text{train}}_{3}\rightarrow\text{D}^{\text{test}}_{4}$ &71.0\% &79.0\% &83.9\% &81.2\% &89.5\% &86.2\% \\
		&$\text{D}^{\text{train}}_{1, 2}\rightarrow\text{D}^{\text{test}}_{4}$ &72.9\% &81.5\% &86.2\% &84.4\% &91.0\% &79.9\%  \\
		&$\text{D}^{\text{train}}_{1, 3}\rightarrow\text{D}^{\text{test}}_{4}$ &67.8\% &80.9\% &85.6\% &82.9\% &91.3\% &82.2\%  \\
		&$\text{D}^{\text{train}}_{2, 3}\rightarrow\text{D}^{\text{test}}_{4}$ &72.6\% &82.8\% &87.2\% &83.3\% &88.7\% &84.6\%  \\
		&$\text{D}^{\text{train}}_{1, 2, 3}\rightarrow\text{D}^{\text{test}}_{4}$ &70.9\% &80.0\% &84.8\% &84.1\% &90.7\% &79.5\%  \\ \hline
		\multirow{7}{*}{R-squared} 
		&$\text{D}^{\text{train}}_{1}\rightarrow\text{D}^{\text{test}}_{4}$ &0.841 &0.884 &0.777 &0.736 &0.776 &0.807  \\
		&$\text{D}^{\text{train}}_{2}\rightarrow\text{D}^{\text{test}}_{4}$ &0.805 &0.876 &0.808 &0.710 &0.770 &0.794  \\
		&$\text{D}^{\text{train}}_{3}\rightarrow\text{D}^{\text{test}}_{4}$ &0.769 &0.867 &0.763 &0.687 &0.785 &0.824  \\
		&$\text{D}^{\text{train}}_{1, 2}\rightarrow\text{D}^{\text{test}}_{4}$ &0.843 &0.882 &0.797 &0.731 &0.776 &0.819  \\
		&$\text{D}^{\text{train}}_{1, 3}\rightarrow\text{D}^{\text{test}}_{4}$ &0.832 &0.881 &0.781 &0.733 &0.782 &0.834  \\
		&$\text{D}^{\text{train}}_{2, 3}\rightarrow\text{D}^{\text{test}}_{4}$ &0.835 &0.894 &0.788 &0.714 &0.780 &0.821  \\
		&$\text{D}^{\text{train}}_{1, 2, 3}\rightarrow\text{D}^{\text{test}}_{4}$ &0.830 &0.875 &0.796 &0.738 &0.782 &0.822  \\ \hline
		\multirow{7}{*}{PPMCC} 
		&$\text{D}^{\text{train}}_{1}\rightarrow\text{D}^{\text{test}}_{4}$ &0.936 &0.943 &0.888 &0.859 &0.881 &0.902  \\
		&$\text{D}^{\text{train}}_{2}\rightarrow\text{D}^{\text{test}}_{4}$ &0.927 &0.939 &0.904 &0.862 &0.882 &0.895  \\
		&$\text{D}^{\text{train}}_{3}\rightarrow\text{D}^{\text{test}}_{4}$ &0.896 &0.935 &0.877 &0.834 &0.889 &0.911  \\
		&$\text{D}^{\text{train}}_{1, 2}\rightarrow\text{D}^{\text{test}}_{4}$ &0.937 &0.941 &0.897 &0.858 &0.882 &0.907  \\
		&$\text{D}^{\text{train}}_{1, 3}\rightarrow\text{D}^{\text{test}}_{4}$ &0.934 &0.942 &0.891 &0.861 &0.885 &0.915  \\
		&$\text{D}^{\text{train}}_{2, 3}\rightarrow\text{D}^{\text{test}}_{4}$ &0.928 &0.946 &0.889 &0.853 &0.888 &0.909  \\
		&$\text{D}^{\text{train}}_{1, 2, 3}\rightarrow\text{D}^{\text{test}}_{4}$ &0.933 &0.940 &0.899 &0.863 &0.885 &0.909  \\ \hline
		\multirow{7}{*}{Outlier-difference} 
		&$\text{D}^{\text{train}}_{1}\rightarrow\text{D}^{\text{test}}_{4}$ &19651 &22317 &16592 &12950 &21951 &14265  \\
		&$\text{D}^{\text{train}}_{2}\rightarrow\text{D}^{\text{test}}_{4}$ &19562 &22361 &16772 &12988 &21959 &13705  \\
		&$\text{D}^{\text{train}}_{3}\rightarrow\text{D}^{\text{test}}_{4}$ &19647 &22125 &16731 &12956 &21931 &15124  \\
		&$\text{D}^{\text{train}}_{1, 2}\rightarrow\text{D}^{\text{test}}_{4}$ &19622 &22333 &16645 &12922 &21955 &14305  \\
		&$\text{D}^{\text{train}}_{1, 3}\rightarrow\text{D}^{\text{test}}_{4}$ &19650 &22277 &16573 &12967 &21961 &13425  \\
		&$\text{D}^{\text{train}}_{2, 3}\rightarrow\text{D}^{\text{test}}_{4}$ &19691 &22072 &16668 &13004 &21949 &15841  \\
		&$\text{D}^{\text{train}}_{1, 2, 3}\rightarrow\text{D}^{\text{test}}_{4}$ &19660 &22313 &16594 &12970 &21954 &13645  \\ \hline
	\end{tabular}
\end{table}

\clearpage

\bibliographystyle{aasjournal}

\bibliography{reference}

\begin{thebibliography}{}
\expandafter\ifx\csname natexlab\endcsname\relax\def\natexlab#1{#1}\fi
\providecommand{\url}[1]{\href{#1}{#1}}
\providecommand{\dodoi}[1]{doi:~\href{http://doi.org/#1}{\nolinkurl{#1}}}
\providecommand{\doeprint}[1]{\href{http://ascl.net/#1}{\nolinkurl{http://ascl.net/#1}}}
\providecommand{\doarXiv}[1]{\href{https://arxiv.org/abs/#1}{\nolinkurl{https://arxiv.org/abs/#1}}}

\bibitem[{Asensio~Ramos \& de~la Cruz~Rodr{\'{\i}}guez(2015)}]{Asensio2015}
Asensio~Ramos, A., \& de~la Cruz~Rodr{\'{\i}}guez, J. 2015, \aap, 577, A140,
  \dodoi{10.1051/0004-6361/201425508}

\bibitem[{Asensio~Ramos \& D{\'{i}}az~Baso(2019)}]{Asensio2019}
Asensio~Ramos, A., \& D{\'{i}}az~Baso, C.~J. 2019, \aap, 626, A102,
  \dodoi{10.1051/0004-6361/201935628}

\bibitem[{Auer {et~al.}(1977)Auer, Heasley, \& House}]{Auer1977}
Auer, L.~H., Heasley, J.~N., \& House, L.~L. 1977, \solphys, 55, 47,
  \dodoi{10.1007/bf00150873}

\bibitem[{Borrero {et~al.}(2011)Borrero, Tomczyk, Kubo, Socas-Navarro, Schou,
  Couvidat, \& Bogart}]{Borrero2010}
Borrero, J.~M., Tomczyk, S., Kubo, M., {et~al.} 2011, \solphys, 273, 267,
  \dodoi{10.1007/s11207-010-9515-6}

\bibitem[{{Cao} {et~al.}(2012){Cao}, {Goode}, {Ahn}, {Gorceix}, {Schmidt}, \&
  {Lin}}]{2012ASPC..463..291C}
{Cao}, W., {Goode}, P.~R., {Ahn}, K., {et~al.} 2012, Astronomical Society of
  the Pacific Conference Series, Vol. 463, {NIRIS: The Second Generation
  Near-Infrared Imaging Spectro-polarimeter for the 1.6 Meter New Solar
  Telescope}, ed. T.~R. {Rimmele}, A.~{Tritschler}, F.~{W{\"o}ger},
  M.~{Collados Vera}, H.~{Socas-Navarro}, R.~{Schlichenmaier}, M.~{Carlsson},
  T.~{Berger}, A.~{Cadavid}, P.~R. {Gilbert}, P.~R. {Goode}, \&
  M.~{Kn{\"o}lker}, 291

\bibitem[{Cao {et~al.}(2006)Cao, Jing, Ma, Xu, Wang, \& Goode}]{Cao2006}
Cao, W., Jing, J., Ma, J., {et~al.} 2006, \pasp, 118, 838,
  \dodoi{10.1086/505408}

\bibitem[{Carroll \& Kopf(2008)}]{Carroll2008}
Carroll, T.~A., \& Kopf, M. 2008, \aap, 481, L37,
  \dodoi{10.1051/0004-6361:20079197}

\bibitem[{Carroll \& Staude(2001)}]{Carroll2001}
Carroll, T.~A., \& Staude, J. 2001, \aap, 378, 316,
  \dodoi{10.1051/0004-6361:20011167}

\bibitem[{Collados(2008)}]{Collados2008}
Collados, M. 2008, in Ground-based and Airborne Telescopes {II}, ed. L.~M.
  Stepp \& R.~Gilmozzi ({SPIE}).
\newblock \url{https://doi.org/10.1117%2F12.790105}

\bibitem[{del Toro~Iniesta \& Ruiz~Cobo(1996)}]{Del1996}
del Toro~Iniesta, J.~C., \& Ruiz~Cobo, B. 1996, \solphys, 164, 169,
  \dodoi{10.1007/bf00146631}

\bibitem[{{Frutiger} {et~al.}(2000){Frutiger}, {Solanki}, {Fligge}, \&
  {Bruls}}]{Frutiger2000}
{Frutiger}, C., {Solanki}, S.~K., {Fligge}, M., \& {Bruls}, J.~H.~M.~J. 2000,
  \aap, 358, 1109

\bibitem[{Galton(1886)}]{Galton1886}
Galton, F. 1886, The Journal of the Anthropological Institute of Great Britain
  and Ireland, 15, 246, \dodoi{10.2307/2841583}

\bibitem[{{Goode} \& {Cao}(2012)}]{2012ASPC..463..357G}
{Goode}, P.~R., \& {Cao}, W. 2012, Astronomical Society of the Pacific
  Conference Series, Vol. 463, {The 1.6 m Off-Axis New Solar Telescope (NST) in
  Big Bear}, ed. T.~R. {Rimmele}, A.~{Tritschler}, F.~{W{\"o}ger}, M.~{Collados
  Vera}, H.~{Socas-Navarro}, R.~{Schlichenmaier}, M.~{Carlsson}, T.~{Berger},
  A.~{Cadavid}, P.~R. {Gilbert}, P.~R. {Goode}, \& M.~{Kn{\"o}lker}, 357

\bibitem[{Goodfellow {et~al.}(2016)Goodfellow, Bengio, \&
  Courville}]{DBLP:books/daglib/0040158}
Goodfellow, I.~J., Bengio, Y., \& Courville, A.~C. 2016, Deep Learning,
  Adaptive computation and machine learning ({MIT} Press).
\newblock \url{http://www.deeplearningbook.org/}

\bibitem[{Lagg {et~al.}(2004)Lagg, Woch, Krupp, \& Solanki}]{Lagg2004}
Lagg, A., Woch, J., Krupp, N., \& Solanki, S.~K. 2004, \aap, 414, 1109,
  \dodoi{10.1051/0004-6361:20031643}

\bibitem[{Landi~Degl'Innocenti(1984)}]{Degl'Innocenti2004}
Landi~Degl'Innocenti, E. 1984, \solphys, 91, 1, \dodoi{10.1007/bf00213606}

\bibitem[{Landolfi {et~al.}(1984)Landolfi, Landi~Degl'innocenti, \&
  Arena}]{Landolfi1984}
Landolfi, M., Landi~Degl'innocenti, E., \& Arena, P. 1984, \solphys, 93, 269,
  \dodoi{10.1007/bf02270839}

\bibitem[{LeCun {et~al.}(2015)LeCun, Bengio, \& Hinton}]{LeCun2015}
LeCun, Y., Bengio, Y., \& Hinton, G. 2015, \nat, 521, 436,
  \dodoi{10.1038/nature14539}

\bibitem[{Liu {et~al.}(2018)Liu, Cao, Chae, Ahn, Choudhary, Lee, Liu, Deng,
  Wang, \& Wang}]{Liu2018}
Liu, C., Cao, W., Chae, J., {et~al.} 2018, \apj, 869, 21,
  \dodoi{10.3847/1538-4357/aaecd0}

\bibitem[{McHugh(2012)}]{McHugh2012}
McHugh, M.~L. 2012, Biochemia Medica, 22, 276, \dodoi{10.11613/bm.2012.031}

\bibitem[{McMullin {et~al.}(2012)McMullin, Rimmele, Keil, Warner, Barden,
  Bulau, Craig, Goodrich, Hansen, Hegwer, Hubbard, McBride, Shimko, Wöger, \&
  Ditsler}]{McMullin2012}
McMullin, J.~P., Rimmele, T.~R., Keil, S.~L., {et~al.} 2012, in Ground-based
  and Airborne Telescopes {IV}, ed. L.~M. Stepp, R.~Gilmozzi, \& H.~J. Hall
  (SPIE).
\newblock \url{https://doi.org/10.1117%2F12.926949}

\bibitem[{Pearson(1895)}]{Pearson1895}
Pearson, K. 1895, Proceedings of the Royal Society of London, 58, 240,
  \dodoi{10.1098/rspl.1895.0041}

\bibitem[{Press {et~al.}(1991)Press, Flannery, Teukolsky, \&
  Vetterling}]{Press1989}
Press, W.~H., Flannery, B.~P., Teukolsky, S.~A., \& Vetterling, W.~T. 1991,
  Mathematics of Computation, 56, 396, \dodoi{10.2307/2008560}

\bibitem[{Quintero~Noda {et~al.}(2015)Quintero~Noda, Asensio~Ramos,
  Orozco~Su{\'{a}}rez, \& Ruiz~Cobo}]{Quintero2015}
Quintero~Noda, C., Asensio~Ramos, A., Orozco~Su{\'{a}}rez, D., \& Ruiz~Cobo, B.
  2015, \aap, 579, A3, \dodoi{10.1051/0004-6361/201425414}

\bibitem[{Rees {et~al.}(2004)Rees, Guo, L{\'{o}}pez~Ariste, \&
  Graham}]{Rees2004}
Rees, D., Guo, Y., L{\'{o}}pez~Ariste, A., \& Graham, J. 2004, in Lecture Notes
  in Computer Science (Springer Berlin Heidelberg), 388--394.
\newblock \url{https://doi.org/10.1007%2F978-3-540-30134-9_52}

\bibitem[{Ruiz~Cobo \& Asensio~Ramos(2012)}]{Ruiz2013}
Ruiz~Cobo, B., \& Asensio~Ramos, A. 2012, \aap, 549, L4,
  \dodoi{10.1051/0004-6361/201220373}

\bibitem[{Ruiz~Cobo \& del Toro~Iniesta(1992)}]{Ruiz1992}
Ruiz~Cobo, B., \& del Toro~Iniesta, J.~C. 1992, \apj, 398, 375,
  \dodoi{10.1086/171862}

\bibitem[{Sen \& Srivastava(1990)}]{SS-1990}
Sen, A., \& Srivastava, M. 1990, Regression Analysis (Springer-Verlag New
  York), \dodoi{10.1007/978-1-4612-4470-7}

\bibitem[{Skumanich \& Lites(1987)}]{Skumanich1987}
Skumanich, A., \& Lites, B.~W. 1987, \apj, 322, 473, \dodoi{10.1086/165743}

\bibitem[{Socas-Navarro(2003)}]{Socas-Navarro2003}
Socas-Navarro, H. 2003, Neural Networks, 16, 355,
  \dodoi{10.1016/s0893-6080(03)00024-8}

\bibitem[{Socas-Navarro(2005)}]{Socas-Navarro2005}
---. 2005, \apj, 621, 545, \dodoi{10.1086/427431}

\bibitem[{Socas-Navarro {et~al.}(2001)Socas-Navarro, L{\'{o}}pez~Ariste, \&
  Lites}]{Socas-Navarro2001}
Socas-Navarro, H., L{\'{o}}pez~Ariste, A., \& Lites, B.~W. 2001, \apj, 553,
  949, \dodoi{10.1086/320984}

\bibitem[{Teng(2015)}]{Teng2015}
Teng, F. 2015, \solphys, 290, 2693, \dodoi{10.1007/s11207-015-0781-1}

\bibitem[{Unno(1956)}]{Unno1956}
Unno, W. 1956, Astronomical Society of Japan, 8, 108

\bibitem[{Wang {et~al.}(2015)Wang, Cao, Liu, Xu, Liu, Zeng, Chae, \&
  Ji}]{Wang2015}
Wang, H., Cao, W., Liu, C., {et~al.} 2015, Nature Communications, 6,
  \dodoi{10.1038/ncomms8008}

\bibitem[{Wang {et~al.}(2017)Wang, Liu, Ahn, Xu, Jing, Deng, Huang, Liu,
  Kusano, Fleishman, Gary, \& Cao}]{Wang2017}
Wang, H., Liu, C., Ahn, K., {et~al.} 2017, Nature Astronomy, 1,
  \dodoi{10.1038/s41550-017-0085}

\bibitem[{Xu {et~al.}(2016)Xu, Cao, Ding, Kleint, Su, Liu, Ji, Chae, Jing, Cho,
  Cho, Gary, \& Wang}]{Xu2016}
Xu, Y., Cao, W., Ding, M., {et~al.} 2016, \apj, 819, 89,
  \dodoi{10.3847/0004-637x/819/2/89}

\bibitem[{Xu {et~al.}(2018)Xu, Cao, Ahn, Jing, Liu, Chae, Huang, Deng, Gary, \&
  Wang}]{Xu2018}
Xu, Y., Cao, W., Ahn, K., {et~al.} 2018, Nature Communications, 9,
  \dodoi{10.1038/s41467-017-02509-w}

\end{thebibliography}

\end{document}